\documentclass[aps,prb,reprint,superscriptaddress]{revtex4-1}

\usepackage{amsmath}
\usepackage{graphicx}
\usepackage{cleveref}

\begin{document}

\title{Epitaxial Stabilisation of ${\bf \mathrm{Ge_{1-x}Sn_x}}$ Alloys}

\author{Alfonso Sanchez-Soares}
\affiliation{EOLAS Designs, Grenagh, Co. Cork, T23 AK70, Ireland}
\affiliation{Tyndall National Institute, University College Cork, Dyke Parade, Cork, T12 R5CP, Ireland}
\author{Conor O'Donnell}
\affiliation{Tyndall National Institute, University College Cork, Dyke Parade, Cork, T12 R5CP, Ireland}
\author{James C.~Greer}%
\affiliation{Department of Electrical and Electronic Engineering and Nottingham Ningbo New Materials Institute, University of Nottingham Ningbo China, 199 Taikang East Road, Ningbo, 315100, China}

\email{Jim.Greer@nottingham.edu.cn}
                             
\begin{abstract} 
The thermodynamic stability of germanium tin $\mathrm{Ge_{1-x}Sn_x}$ alloys is investigated across the composition range $0 \le x \le 1$ by applying density functional theory (DFT) together with the cluster expansion formalism (CE). It is known that GeSn alloys are immiscible and that non-equilibrium growth techniques are required to produce metastable films and nanostructures. Insight into the driving forces behind component segregation is gained by investigating the equilibrium thermodynamics of GeSn systems. The alloy free energy of mixing is computed by combining enthalpies from CE with entropy terms for configurational and vibrational degrees of freedom. Volume deformations due to the large mismatch in ionic radii are readily found to be the key driving force for immiscibility at all temperatures of relevance. This leads to a study of epitaxial stabilisation by employing latticed matched substrates to favour growth of alloys with fractional compositions of $\mathrm{x=0}$, approximately $\mathrm{x=0.5}$ and $\mathrm{x=1}$. Reduction of the free energy of mixing due to epitaxial strain in thin films is quantified for each substrate leading to indicators for growth of kinetically stable films.
\end{abstract}

\maketitle

\section{\label{sec:Intro}Introduction}

Germanium tin (GeSn) alloys have received significant attention over the past decade due to their potential applications in optoelectronic devices.\cite{Doherty2020} At low tin content these alloys have been shown to exhibit a direct energy band gap, making them particularly suitable for photonics devices that promise the possibility of being integrated into silicon-based technologies, as GeSn has been grown employing CMOS-compatible processes.\cite{Wang2007,Conley2014} Although GeSn alloys with low Sn concentrations have been reported in the literature for some time, progress towards general technological applications has been much slower mainly due to two of its key characteristics: i) Ge and Sn atoms exhibit a large difference in ionic radii which results in a solid solubility of tin in germanium of less than $ 1 \%$,\cite{OlesinskiAbbaschian1984} and ii) tin's stable phase under standard conditions is the tetragonal $\beta$-Sn structure instead of the preferable cubic $\alpha$-Sn structure, which only becomes stable at temperatures below $13^{\circ} $C at standard pressure.~\cite{Raynor1958a} These characteristics result in significant difficulty in incorporating Sn into a cubic Ge lattice without the precipitation of metallic $\beta$-Sn clusters. Significant progress has been made in recent years towards fabricating GeSn alloys by employing non-equilibrium epitaxial growth techniques such as molecular beam epitaxy (MBE) and chemical vapour deposition (CVD), allowing realization of thin films with thicknesses up to hundreds of nanometres.~\cite{Piao1990,Gencarelli2013,Conley2014,WirthsBucaMantl2016,Dou2018, WEI2021125996, Dascalescu2020, Tai2020} Moreover, recent studies have reported lasers comprised of $\mathrm{Ge_{1-x}Sn_{x}}$ thin films with low tin content ($0.054 < \mathrm{x} < 0.16$), indicating significant advances towards fabrication of alloys with device-grade crystallinity.~\cite{Wirths2015,Reboud2017,Rainko2019, Elbaz2020, Chretien2019}  \\

Although recent efforts have been focused on low tin content alloys with a direct band gap for applications in photonics, the electronic structure of germanium-tin alloys has been predicted to vary significantly with composition: while the addition of a few atomic percent of tin into germanium induces a transition from indirect to a direct band gap, alloys with larger tin contents exhibit semimetallic behaviour.~\cite{Polak2017,ODonnell2020} Integration of direct band gap $\mathrm{Ge_{1-x}Sn_{x}}$ into electronic devices promises advantages such as enhanced carrier mobilities, optical interconnects, and improved efficiency in tunnelling-based devices such as tunnelling field-effect transistors (TFETs) or tunnel diodes.~\cite{Sau2007, Kao2012, Schulte-Braucks2015} However, the possibility of increasing tin composition to induce semimetallic regions or to counteract confinement-induced band gap widening in nanoscale devices further increases the technological value of these alloys. With critical dimensions for modern electronics already below $10$ nm, designs capable of exploiting semimetals using confinement- and surface-assisted band gap engineering are being considered, and the ability to further modify band gaps by varying alloy composition introduces another parameter to achieve new nanoelectronic device designs.~\cite{Ansari2012,Sanchez-Soares2016,Ansari2016,Sanchez-Soares2016a,Gity2017,Gity2018} The growth of crystalline $\mathrm{Ge_{1-x}Sn_{x}}$ alloys with intermediate tin contents has also been experimentally achieved; recently, atomically flat epitaxial films grown on $\mathrm{Ge}(100)$ with tin content as high as $\mathrm{x=0.46}$ were reported in the literature.~\cite{Suzuki2016} In their study, compressive strain due to high Sn content of films pseudomorphically grown on germanium substrates resulted in crystalline films with a critical thickness of 3 nm. In contrast, the use of lattice-matched substrates has been reported to allow growth of alloys with compositions $\mathrm{0.26 < x < 0.99}$ and even stabilisation of pure $\alpha$-Sn films with thicknesses in excess of 100 nm.~\cite{Farrow1981,Hoechst1983,Asom1989,John1989,Bowman1990,Piao1990} Strain reduction by growth on lattice-matched substrates thus allows fabrication of semimetallic $\mathrm{Ge_{1-x}Sn_{x}}$ with thicknesses on the order of hundreds of nanometres, which is well above the lengths required for nanoelectronic applications.\\

This work investigates the thermodynamic stability of bulk and epitaxial $\mathrm{Ge_{1-x}Sn_{x}}$ alloys. Density functional theory (DFT) calculations are performed in conjunction with the cluster expansion (CE) formalism to quantify the magnitude of various contributions to the Helmholtz free energy of mixing that lead to immiscibility in bulk GeSn alloys. The case of epitaxially grown alloys is then explored by analysing the impact of constraining the lattice constant of alloys along a plane on the structural properties and free energy contributions. Results suggest that high tin content alloys may be epitaxially grown by appropriate choice of substrate material and orientation.

\section{Methods}\label{sec:Method}

Computational methods are applied to study the thermodynamic stability of $\mathrm{ Ge_{1-x} Sn_x }$ alloys for fractional compositions $0\le \mathrm{x} \le 1$. The Helmholtz free energy is given by
\begin{equation}\label{eq:helmholtz}
F(\mathrm{x}) = E(\mathrm{x}) - T S(\mathrm{x})
\end{equation}
with $E$ the internal energy, $T$ the temperature, and $S$ the entropy; electronic, vibrational and configurational contributions to the free energy are considered. The configurational contribution to the internal energy is considered using two approximations: the Bragg-Williams (BW) model and the cluster expansion (CE) formalism.~\cite{Bragg1934,WilliamsII1935, Williams1935,Sanchez1984} Both methods define the configurational energy in terms of interaction parameters that can be extracted from density functional theory (DFT) calculations using periodic supercells. In the BW model, the bond energies do not account for local variations in the alloy composition, whereas the CE is designed to describe all bonding configurations arising in a random alloy. The DFT calculations applied here to determine the alloy interaction energies employ the usual Kohn-Sham (KS) framework in an implementation using norm-conserving pseudopotentials and linear combination of numerical atomic orbitals (NAO) basis sets.\cite{Soler2002,Ozaki03,Ozaki04,QW} Pseudopotentials include 4s and 4p electrons to describe germanium and 4d, 5s, and 5p electrons for tin. Basis sets used to expand wavefunctions include s4p4d3f2 NAO for germanium and s2p3d3f2 NAO for tin, where the notation indicates the number of $s$-type, $p$-type, $d$-type, and $f$-type orbitals centred about atoms of each species. Brillouin zone integrations are performed on a $k$-point grid generated according to the Monkhorst-Pack scheme with a density of at least 7 $k$-points / \rm{\AA}$^{-1}$.~\cite{Monkhorst1976} Real-space quantities are discretised on a grid with a minimum cut-off energy of 100 Hartree. A parametrisation of the local density approximation (LDA) is employed for the exchange-correlation potential.~\cite{Perdew1981} \\

Minimisation of the total energy with respect to ionic positions and periodic cell dimensions (geometry optimization) is performed on all supercells to determine relaxed geometries, which are defined as having a maximum force acting on an atom less than a threshold of $\vert\, 5 \times 10^{-2}\, \mathrm{eV/}$\rm{\AA}$\,\vert$ and with all elements of the corresponding stress tensors less than $\vert\, 0.1 ~\mathrm{GPa}\,\vert$. The inclusion of spin-orbit interactions in the calculations is found to have a relatively small impact on the final atomic positions and total energies for a reference set of alloy supercells and thus is neglected, a finding consistent with a recent study for germanium tin alloys.~\cite{Polak2017} However, it is noted that spin-orbit coupling can have a significant influence on KS band dispersions.~\cite{ODonnell2020} \\

The structural properties of tin and germanium in the diamond structure computed with this approach are given in \cref{tbl:bulk_lat}. The calculated equilibrium lattice parameters show agreement with experimentally reported values to within 1\%.~\cite{Farrow1981,Baker1975,Thewlis1954} The bulk modulus as calculated for germanium exhibits a deviation of less than 3\% with respect to the experimental values obtained from ultrasound techniques, and the calculated bulk modulus for tin in the $\alpha$-phase lies within previously reported values obtained from neutron scattering experiments.~\cite{Bruner1961,Price1971,Buchenauer1971} \\

The BW model is a mean-field model resulting from a nearest neighbour description for the configurational energy of a substitutional alloy
\begin{equation} \label{eqn:Bragg_U}
E_{\mathrm{config}} = \sum_{i,j= \lbrace\mathrm{Ge,Sn}\rbrace} V_{ij} N_{ij},
\end{equation}
with $V_{ij}$ and $N_{ij}$ giving the bond energies and number of bonds, respectively, between atoms of type $i$ and $j$ located on lattice sites. The BW model describes the atom-atom interactions in GeSn alloys solely in terms of the three possible nearest neighbour interaction or bond energies $V_{\rm GeGe}, V_{\rm GeSn}=V_{\rm SnGe}$, and $V_{\rm SnSn}$ yielding an expression as
\begin{align}
E_{\mathrm{config}}(\mathrm{x}) 
=&\frac{Nz}{2}\big[ (1-\mathrm{x})^2 V_{\mathrm{GeGe}}
+\mathrm{x}^2   V_{\mathrm{SnSn}} \nonumber \\  
&+ 2 \mathrm{x} (1-\mathrm{x}) V_{\rm GeSn} \big],
\end{align}
where $z$ is the nearest neighbour coordination number and $N$ the number of atoms in a considered cell. Germanium tin alloys under typical experimental growth conditions are random alloys.~\cite{Biswas2016,Mukherjee2017} Topologically, each lattice site is given an occupation Ge or Sn with probability ${\rm (1-x)}$ and ${\rm x}$, respectively, with each lattice site having four nearest neighbours as per bonding in non-amorphous group IV materials. To determine the nearest neighbour interactions in the BW model, energies for the homonuclear bonds Ge-Ge and Sn-Sn are obtained from the DFT total energies for relaxed diamond cells and for isolated Ge and Sn atoms, respectively. For the heteronuclear bond energy, a DFT total energy calculation is performed for a fictitious zinc-blende germanium tin alloy and taking the aforementioned isolated atom simulations as reference. This parametrisation neglects bond energy changes due to local variations in the occupation of nearest neighbour atom sites and from higher coordination shells. However, the BW model yields a useful indicator of alloy stability through the relative binding energy 
\begin{equation}
V=V_{\mathrm{GeSn}} - \frac{1}{2}(V_{\mathrm{GeGe}} + V_{\mathrm{SnSn}} ),
\end{equation}
which provides a comparison of the energy of a heteronuclear bond to the average energy of the homonuclear bonds. The magnitude and sign of the relative binding energy provide a clear and intuitive indication for a preference to segregate or to form a solid solution. \\

The most severe limitations inherent to the BW model can be overcome by including interactions beyond nearest neighbours through the cluster expansion formalism. By characterizing lattice site occupations $i$ using spin variables corresponding to the atom type as $\sigma_i$, analogous to the Ising model, the configuration within a supercell representing an alloy can be represented by a vector $\sigma = \{\sigma_i\}$. The internal configurational energy may then expanded as
\begin{equation}
E_{\mathrm{config}}(\sigma) = \sum_\alpha J_\alpha \hat{S}_\alpha (\sigma),\\
\label{eq:ce}
\end{equation}
whereby lattice sites are grouped into clusters denoted $\alpha$ with associated pseudospin $\hat{S}_{\alpha}$, which is the average product of lattice site \emph{spins} over each set of unique clusters, and $J_{\alpha}$ are effective cluster interactions (ECI). By including all clusters, the internal energy of any alloy configuration can be exactly reproduced. However the main effects of the chemical bonding are generally speaking localized and the magnitudes for the ECIs are found to rapidly decrease with increasing cluster size and range, thus the internal energy of an alloy can be well approximated by truncating the sum in \cref{eq:ce} to include relatively small and short range clusters. The ECI in the truncated sum can then be extracted from explicit total energy calculations performed on a relatively small set of configurations in an approach known as the Connolly-Williams method or the structure inversion method.~\cite{Connolly1983} In order to define an optimal set of clusters to include when building the CE for the germanium tin alloys, a set of 47 crystalline test structures with a maximum of 8 atoms per supercell are generated following an algorithm designed to maximise the predictive power of the fit whilst minimising the computational effort of the required first principles calculations.~\citep{Walle2002} The total energies for the test structures are used to extract the ECI coefficients for two atom clusters with a maximum range of 1 nm and for three atom clusters with a maximum range of 0.52 nm. The selection of clusters to include in the CE applies a statistical measure relying on a cross-validation score and variance minimisation techniques to reproduce the supercell energies for the test structures.~\cite{Stone1974} The resulting cross validation score is 5 meV/atom, which is comparable to the error expected from the DFT energies used to determine the ECI values. \\

\begin{table}
\caption{Structural parameters of germanium and $\alpha$-tin calculated with DFT LDA compared to reported experimental values.}
\label{tbl:bulk_lat}
\begin{ruledtabular}
\begin{tabular}{ccccc}

  & \multicolumn{2}{c}{$a_0$ (\AA)} & \multicolumn{2}{c}{$B_0$ (GPa)} \\
         & This work & Exp. & This work & Exp.\\
	\hline 
  Ge		&   $5.64$	& $5.657$\cite{Baker1975}	&$74$ & $75.8$\cite{Bruner1961}\\ 

 Sn			& $6.47$ 	& $6.489$\cite{Farrow1981, Thewlis1954}	&  $47.16$ &$42.5-53.1$\cite{Price1971, Buchenauer1971}\\ 
\end{tabular}
\end{ruledtabular}
\end{table}	

The temperature dependence of the free energy requires determination of the configurational, electronic and vibrational contributions to the entropy. The configurational entropy is estimated in a first approximation by enumerating unique atomic configurations for a random alloy at each composition. Electronic contributions to the free energy are computed from a temperature-independent, one-electron band model employing the density of states calculated from DFT.~\cite{Wolverton1995,Walle2009} These electronic band calculations are performed with a set of supercells designed to mimic the radial correlation functions of random alloys known as special quasirandom structures (SQS);~\cite{Zunger1990, Hass1990} 64-atom cubic supercells ($2 \times 2 \times 2$ simple cubic) have been generated stochastically using a simulated annealing procedure.~\cite{VandeWalle2013} The dimensions of generated SQS allow targeting the site-site correlations of clusters found to be most relevant by the CE whilst maintaining DFT calculations tractable. Vibrational contributions to the free energy are estimated using a bond stiffness versus bond length model. Within this model, force constant tensors are obtained by computing reaction forces on crystalline structures with slight atomic displacements and at varying levels of strain.~\cite{Walle2002a,Fultz2010} Three crystalline structures are chosen: i) elemental Ge; ii) elemental Sn; and again iii) a fictitious GeSn zinc-blende structure. Cubic supercells with 64 atoms are employed such that the minimum distance between displaced atoms is at least 1.2 nm. Reaction forces are computed at 10 different levels of strain and for atomic displacements limited to 0.02 nm away from ideal lattice sites. The force constant tensors are used to perform a polynomial fit relating bond lengths to bond stiffness; this parametrisation can then be used on arbitrarily large supercells to calculate phonon frequencies and determine vibrational contributions to the free energy $F_{vib}$.~\cite{Bozzolo2007} Additional details on the bond stiffness vs. bond length model and comparisons of extracted vibrational properties with experimental literature are provided as Supplementary Information. \\

Two cluster expansions are defined: the cluster expansion of the configurational energy from a set 47 crystalline cells as described previously, and a second expansion for the temperature-dependent vibrational contributions to free energy computed from 64-atom SQS. These two expansions can then be merged to produce a set of temperature dependent ECI. In the following, it will be shown that the electronic contributions to the free energy are negligible, and as such are ignored. Once the TECI are determined, they can be implemented in semi-grand canonical ensemble Monte Carlo simulations to calculate a phase diagram. Configurational entropy contributions are readily included in the MC method as the simulation samples accessible configurations at a given temperature. In the semi-grand canonical ensemble, the number of atoms is fixed, with the energy and composition allowed to fluctuate based on an externally imposed difference in chemical potential between the two atom types as well as an externally imposed temperature.~\cite{Walle2002b} Supercells based on a face-centred cubic primitive cell ($84 \times 84 \times 84$) and with a corresponding lattice constant of 26 nm and 1,185,408 atomic sites are employed for an improved description of long-range order and ensemble averages. The number of equilibration passes and the averaging passes are determined based on a target precision of $10^{-3}$ for the atomic composition of the phase. The phase diagram across the full alloy composition range is then determined using the procedure described in ref.~\onlinecite{Walle2002b}.

\section{\label{sec:Results}Results and Discussion}

\subsection{Bulk alloys} 

The configurational energy of mixing or formation energy is evaluated in a first approximation using the BW model for the composition range $0 \le x \le 1$. Computed nearest neighbour interactions result in a relative binding energy value of $V = +11 ~\mathrm{meV}$: although its positive sign indicates phase segregation is energetically favourable, its magnitude is near the precision limit of DFT simulations employed in its calculation. Thus while the BW model predicts a slight preference for the system to segregate,  kinetically grown, strain-free metastable alloys could exhibit random species distribution due to the relatively large energetic cost of rearranging bond types when compared to predicted stability gains. The magnitude of the relative binding energy also suggests that entropy contributions to the free energy may have a substantial impact on reducing the the free energies for mixing tending to stabilise metastable structures at typical growth temperatures. \\

\begin{figure}
%\centering
\includegraphics[width=\columnwidth]{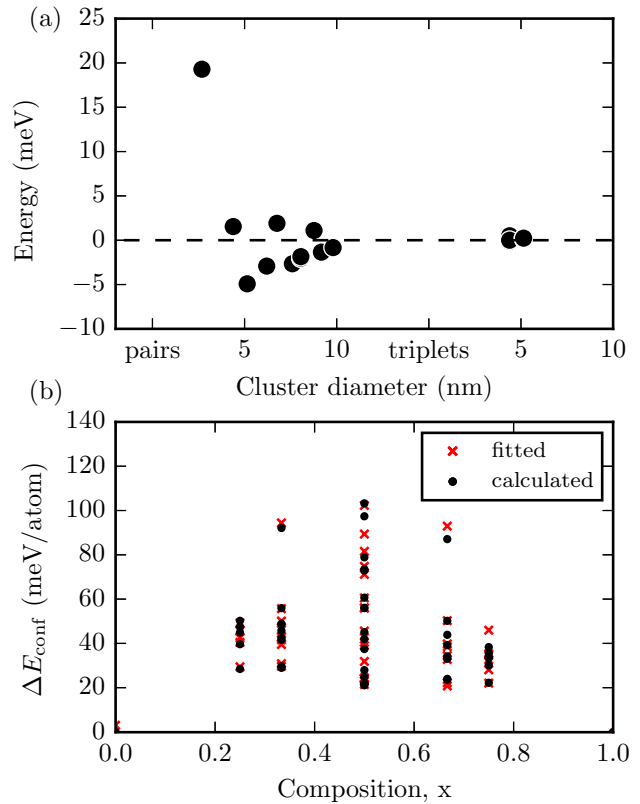}
\caption{(a) Magnitude of effective cluster interactions obtained from the cluster expansion fit for bulk alloys with energies from DFT, and (b) comparison between formation energies predicted by the CE and calculated using DFT for the set of structures employed in the fit.}\label{fig:ce_ranges}
\end{figure}

To provide a more accurate treatment of the free energies and to identify the dominant stabilising effects for germanium tin alloys, the CE is applied to account for configurational variations beyond the nearest neighbour description provided by the BW model.  \Cref{fig:ce_ranges}(a) shows the extracted ECI coefficients for the CE. The magnitude of two-atom cluster ECIs are seen to decay rapidly with distance from a maximum of approximately 20 meV/atom for the nearest-neighbour pair cluster to less than 5 meV/atom for longer range pairs. The ECIs of the three-atom clusters are an order of magnitude smaller than typical values for pair clusters, with a maximum value of 0.5 meV/atom. Consistent with the simple BW analysis and in agreement with experiment, the CE results also predict germanium and tin to be immiscible at zero temperature and thermodynamic equilibrium as shown in \cref{fig:ce_ranges}(b). \\

\begin{figure}
\includegraphics[width = \columnwidth]{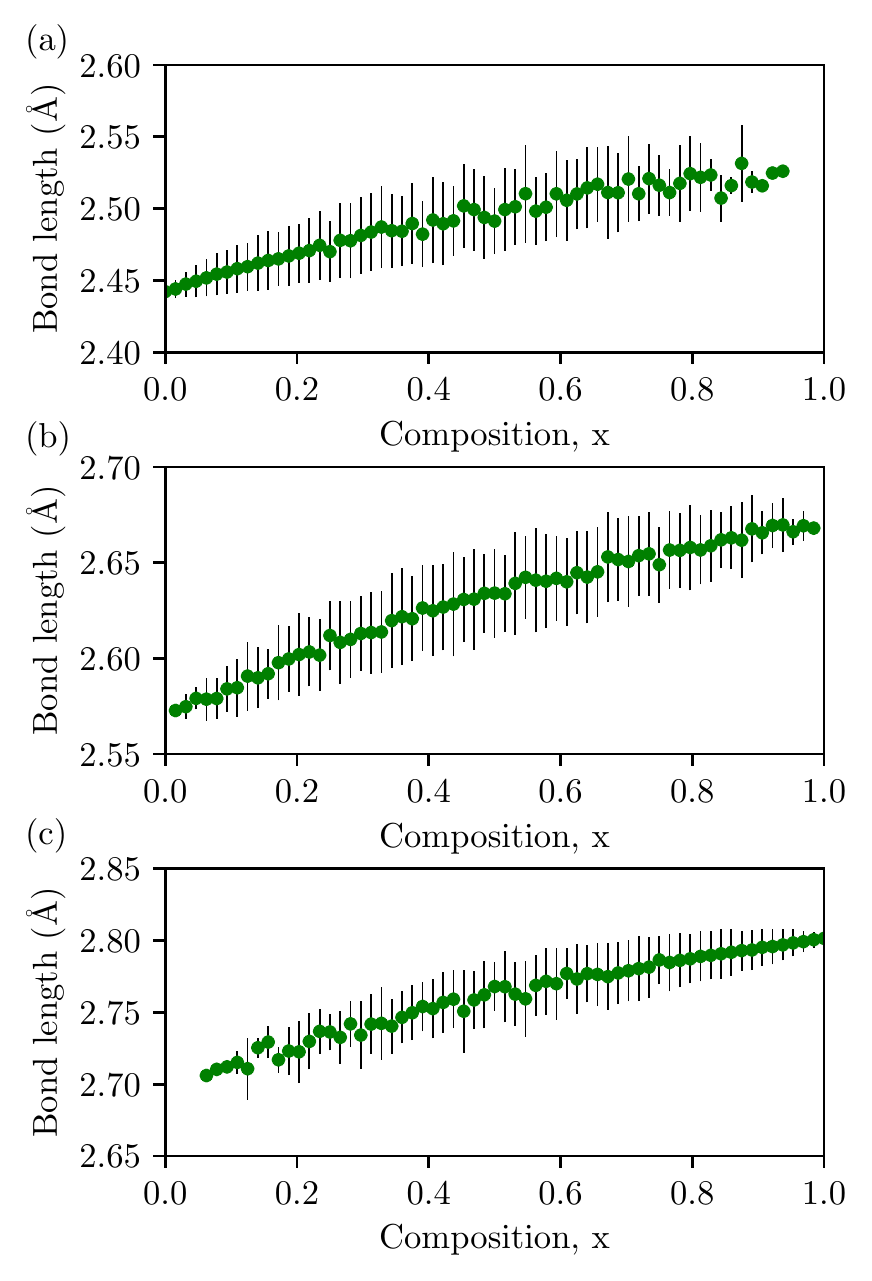} 
\centering
\caption{Average (a) Ge-Ge, (b) Ge-Sn and (c) Sn-Sn bond lengths of $\mathrm{Ge_{1-x}Sn_x}$ from 64-atom SQS relaxed using DFT (green dots), across the full composition range with black bars indicating standard deviations.}
\label{fig:bond_lengths}
\end{figure}

\begin{figure}
\includegraphics[width = \columnwidth]{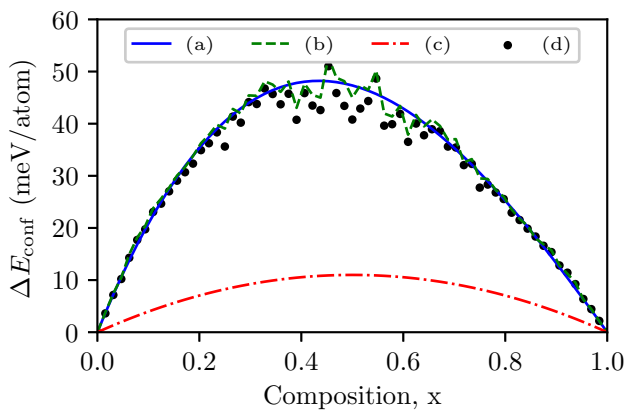} 
\centering
\caption{Formation energy at $T=0$ K as calculated (a) using the CE fit for random alloys, (b) applying the CE fit to generated 64-atom SQS, (c) from the BW model, and (d) directly from DFT calculations for 64-atom SQS.}
\label{fig:SQS_FE}
\end{figure}

Results obtained with lattice models are compared directly to DFT calculations over the set of 64-atom SQS. Atomic positions and lattice constants in the SQS are relaxed using DFT by constraining the cell shape to cubic while allowing the cell volume to vary, an approach previously argued to provide total energies representative of disordered states.~\cite{Ghosh2008} \Cref{fig:bond_lengths} shows the evolution of Ge-Ge, Ge-Sn, and Sn-Sn bond lengths with alloy composition as determined from SQS. Bond lengths calculated for low Sn composition deviate less than 2\% with respect to values reported in previous theoretical and experimental literature.\cite{OHalloran2019,Gencarelli2015} The difference in atomic radii between Ge and Sn and the local variations in chemical environments result in bond lengths with standard deviations on the order of 1\% at intermediate compositions. The alloy lattice constants are compared to literature values by fitting the optimized SQS lattice constants $a_0$ at varying compositions using 

\begin{equation}
a_0^{\mathrm{Ge_{\mathrm{1-\mathrm{x}}}Sn_{\mathrm{x}}}} = a_0^{\mathrm{Ge}} (1 - \mathrm{x}) + a_0^{\mathrm{Sn}}\mathrm{x} + b_a \mathrm{x}(1-\mathrm{x}),
\end{equation}
from which a bowing parameter value of $b_a=0.0056$ nm is obtained, consistent with values reported in other recent theoretical and experimental works.~\cite{Gencarelli2013,Xu2017,Polak2017} \\

\Cref{fig:SQS_FE} compares the formation energy of random alloys across the full composition range as predicted with the BW model, the cluster expansion, and from DFT calculations of SQS. It is observed that although the BW model correctly predicts immiscibility, it underestimates the formation energy of GeSn random alloys by an order of magnitude with respect to results obtained from the CE fit and from 64-atom SQS. Hence there is a stronger tendency for germanium and tin to segregate than the local bond analysis from the BW model indicates. The improved treatment by the CE is seen through its ability to reproduce the energies of relaxed SQS with an RMS error of 2 meV/atom (green dashed curve). \\

\begin{figure}
\centering
\includegraphics[width=\columnwidth]{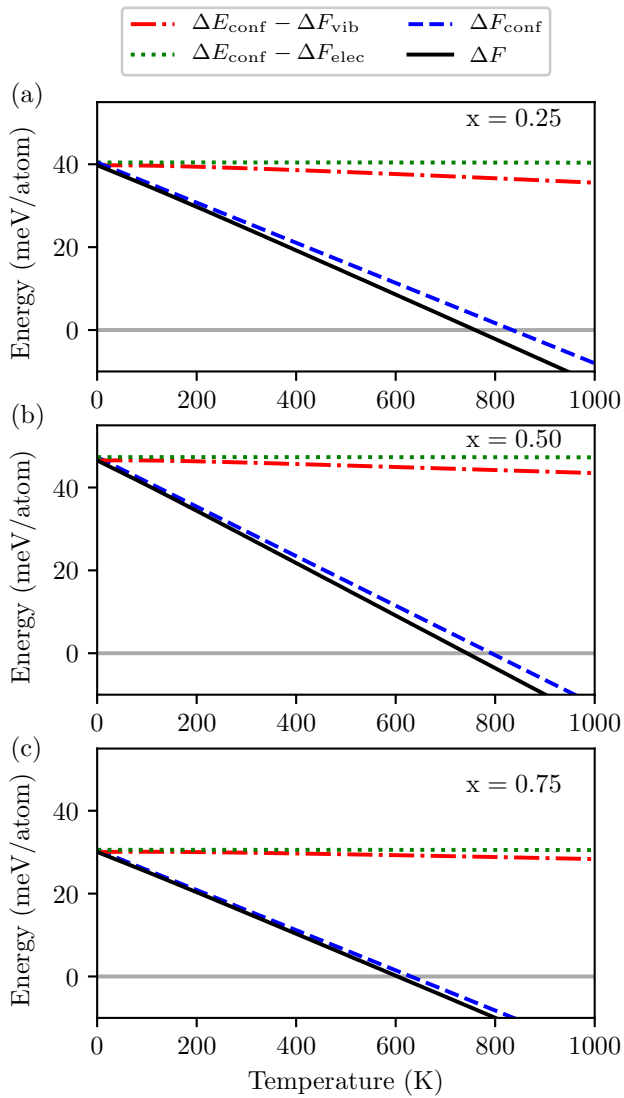}
\caption{Temperature dependence of the total free energy of mixing and considered contributions as estimated from CE fits for (a) $\mathrm{x=0.25}$, (b) $\mathrm{x=0.50}$, and (c) $\mathrm{x=0.75}$ alloy compositions.}
\label{fig:f_vs_t}
\end{figure}

The free energy at finite temperature has been computed by including the electronic, vibrational, and configurational contributions. Electronic contributions are computed from the density of states using the KS band structures obtained for the SQS. The vibrational contributions are calculated by the bond stiffness versus bond length model, and configurational entropy is estimated by enumerating unique configurations at a given composition for a cell containing a large number of atoms, using the Boltzmann entropy expression and Stirling's approximation to yield 

\begin{equation}
S_{\mathrm{conf}}(\mathrm{x}) = -k_B N [ \mathrm{x} \ln \mathrm{x} + (1-\mathrm{x}) \ln (1-\mathrm{x})],    
\end{equation}
with $k_B$ Boltzmann's constant. We note that while this simplified approach assumes all alloy configurations at a given composition are equally accessible and thus represents an upper limit to configurational entropy, it provides estimates useful for comparing the magnitude of various contributions to free energy. \Cref{fig:f_vs_t} shows the resulting temperature dependence of the total free energy of mixing and its various contributions for the representative compositions of $\mathrm{x} = 0.25$, $0.50$, and $0.75$; it is seen that the calculated electronic entropy contributions remain below 1 meV/atom for temperatures up of 1000 K for all three alloy compositions. The configurational entropy contributions dominate the temperature dependence of the free energy of mixing and serve as the main stabilising force at all temperatures of interest, with vibrational contributions favouring alloy formation to a lesser degree. The critical temperature, {\it i.e.} temperatures at which the disordered phase becomes energetically favourable, is seen  to decrease for larger tin compositions in a result attributable to the asymmetry observed in \cref{fig:SQS_FE} for the composition-dependent formation energy $\Delta E_{\mathrm{conf}}$; while this asymmetry is reproduced by both the CE fit and SQS calculations, the formation energy in the BW model reduces to $\Delta E_{\mathrm{conf}} = V[\mathrm{x(1-x)}]$ which is symmetric about $\mathrm{x}=0.5$. \\

\begin{figure}
\centering
\includegraphics[width=\columnwidth]{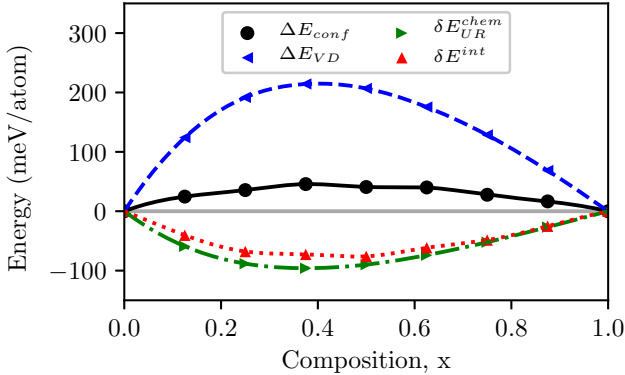} 
\caption{Formation energy decomposition obtained from DFT simulations of bulk alloys employing 64-atom SQS.}
\label{fig:e_decomp}
\end{figure}

To further investigate the origin of the asymmetry, the formation energy is partitioned as
\begin{equation}\label{eq:decomp}
\Delta E_{\mathrm{conf}} = \Delta E_{VD} + \delta E^{\mathrm{chem}}_{UR} + \delta E^{int},
\end{equation}
where the three contributions correspond to the volume deformation energy, the chemical or \emph{spin-flip} energy, and the internal relaxation energy. The volume deformation energy $\Delta E_{VD}$ is defined as the energy required to hydrostatically strain Ge and Sn cells to the equilibrium lattice parameter corresponding to a specific alloy composition. The resulting strain energies are weighted by their respective fractional compositions in the alloy 
\begin{align}\label{eq:decomp2}
\Delta E_{VD} =& (1-\mathrm{x}) \big[E_{\mathrm{Ge}} (a_{\mathrm{Ge_{1-x}Sn_x)}}-E_{\mathrm{Ge}}(a_{\mathrm{Ge}})\big] \nonumber\\
&+\mathrm{x}\big[E_{\mathrm{Sn}}(a_{\mathrm{Ge_{1-x}Sn_x}})-E_{\mathrm{Sn}}(a_{\mathrm{Sn}})\big],
\end{align}
where $E_i(a_j)$ is the energy for $i$= \{Ge,Sn\} at the equilibrium lattice constant $a_j$ for $j$=\{Ge,Sn, Ge$_{1-\mathrm{x}}$Sn$_\mathrm{x}$\}. The chemical or \emph{spin-flip} energy $\delta E^{\mathrm{chem}}_{UR}$ corresponds to the energy gained when germanium and tin atoms occupying (unrelaxed--$UR$) diamond lattice sites form heteronuclear bonds
\begin{align}\label{eq:decomp3}
\delta E^{\mathrm{chem}}_{UR}=& E_{\mathrm{Ge_{1-x}Sn_x}}(a_{\mathrm{Ge_{1-x}Sn_x}}) \nonumber\\&- \big[(1-\mathrm{x})E_{\mathrm{Ge}}(a_{\mathrm{Ge_{1-x}Sn_x}}) + \mathrm{x} E_{\mathrm{Sn}}(a_{\mathrm{Ge_{1-x}Sn_x}})].
\end{align} 
The internal relaxation energy $\delta E^{int}$ is the energy gained when the atomic positions in an alloy are allowed to relax by displacing them from the ideal diamond lattice sites:
\begin{eqnarray}
\label{eq:decomp4}
\delta E^{int} = E_{\mathrm{Ge_{1-x}Sn_x}}(a_{\mathrm{Ge_{1-x}Sn_x}})\vert_R - \\ \nonumber E_{\mathrm{Ge_{1-x}Sn_x}}(a_{\mathrm{\mathrm{Ge_{1-x}Sn_x}}})\vert_{UR},
\end{eqnarray}
where $R$ and $UR$ indicate whether the atomic positions are relaxed or constrained to the ideal diamond lattice sites, respectively. \Cref{fig:e_decomp} shows the dependence of each contribution as computed for 64-atom SQS across the composition range, and \cref{tbl:rel_vol_decomp} lists their magnitude at compositions x=0.25, x=0.50, and x=0.75. The large difference between the Ge and Sn equilibrium lattice parameter results in the volume deformation energy dominating and destabilising the alloys, and the marked asymmetry observed for this quantity can be ascribed to the significant difference between components' bulk moduli; the volume expansion required to accommodate Sn into Ge-rich alloys increases formation energy significantly more than the corresponding volume contraction associated with incorporation of Ge into Sn-rich alloys. Contributions arising from chemical interactions between different atomic species and the relaxation of internal coordinates are of a similar magnitude and act to stabilise the alloys, partially offsetting the effects of the volume deformations. Although the magnitude of stabilising contributions correlates with the magnitude of volume deformation energies, the asymmetry present in the latter dominates formation energies and thus results in larger destabilisation for germanium-rich alloys compared to tin-rich alloys.  \\

\begin{table}
\begin{ruledtabular}
\begin{tabular}{ccccc}

Composition & $\Delta E_{\mathrm{conf}}$ & $\Delta E_{VD} $ & $\delta E^{chem}_{UR}$ & $\delta E^{int}$ \\
(x) & \multicolumn{4}{c}{(meV/atom)} \\
	\hline
0.25 & 36 & 192 & -88 & -67 \\
0.50 & 41 & 207 & -90 & -76 \\
0.75 & 28 & 128 & -52 & -48 \\
\end{tabular}
\end{ruledtabular}
\caption{Formation energy decomposition into contributions listed in \cref{eq:decomp} for bulk alloys at selected compositions computed from 64-atom SQS.}\label{tbl:rel_vol_decomp}
\end{table}	

\begin{figure}
\includegraphics[width=\columnwidth]{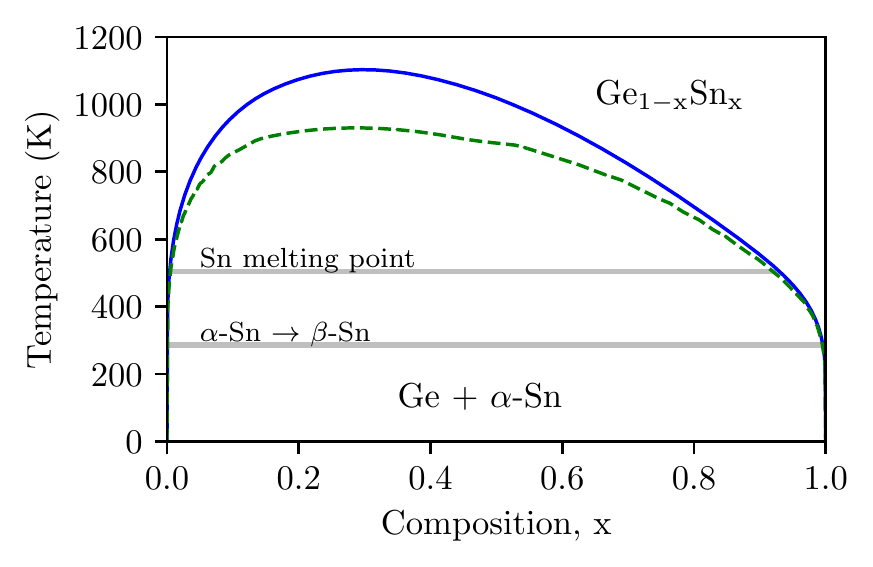} 
\centering
\caption{Phase diagram of bulk alloys as computed using a lattice model Monte Carlo simulation. CE fits constructed including (dashed line) and excluding (solid line) vibrational degrees of freedom have been employed. Regions above the curves represent stability for random alloys. Horizontal lines indicate temperatures at which phase transformation and melting occurs for Sn.}
\label{fig:phase_diagrams}
\end{figure}

The impact of the asymmetry on the system's critical temperature is revealed by the phase boundary between the disordered phase (i.e. random alloy) and decomposition into the elemental phases. Lattice model Monte Carlo simulations employing the CE fit enable larger supercells than tractable with DFT to be studied, resulting in a much improved treatment for ensemble averages over large numbers of alloy configurations. Temperature effects are included by combining ECIs describing the internal configurational energies with configurational and vibrational free energy contributions. The resulting phase diagram for $\mathrm{Ge_{1-x}Sn_x}$ is shown in \cref{fig:phase_diagrams}, where two horizontal lines indicating the temperatures for the $\alpha \to \beta$ phase transition and melting point of tin have been included for reference. The higher transition temperatures found with this method when compared to those shown in \cref{fig:f_vs_t} are attributed to a more accurate treatment of configurational entropy contributions, as the MC simulation accounts for energy variations across alloy configurations. Germanium-rich alloys with compositions up to $x=0.005$ are predicted to be stable below the eutectic temperature, in agreement with previous reports;\cite{OlesinskiAbbaschian1984} for tin-rich alloys stability is only observed for compositions $x>0.995$ at temperatures where $\alpha$-Sn remains stable. Predicted critical temperatures at intermediate compositions are well above the melting temperature for tin. These results clearly indicate that growth of bulk equilibrium $\mathrm{Ge_{1-x}Sn_x}$ alloys at technologically relevant compositions is not possible at standard pressure. It is worth noting that inclusion of vibrational free energy contributions reduces predicted critical temperatures, with larger reductions for germanium-rich compositions. This can be attributed to a stabilising effect associated from softening of phonon modes by addition of tin, partially counteracting the destabilising effects of volume deformations.~\cite{Gassenq2017,Xu2019,Bouthillier2020}

\begin{figure}
\includegraphics[scale=.85]{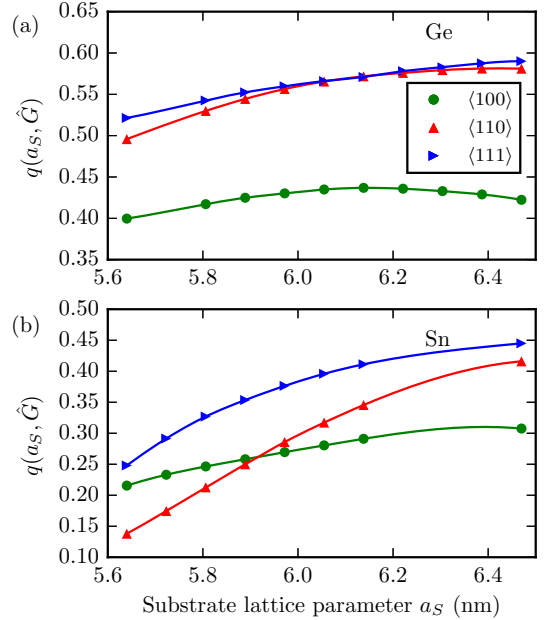}
\centering
\caption{Epitaxial softening functions for (a) Ge and (b) Sn as computed from DFT calculations. For reference, the relaxed lattice constants obtained from DFT are $a_0^\mathrm{Ge}=0.564$ nm and $a_0^\mathrm{Sn}=0.647$ nm}
\label{fig:Epi_soft_funcs}
\end{figure}

\subsection{Epitaxial alloys}

The primary factor leading to the high critical temperatures has been shown to arise from the large difference in equilibrium volumes for germanium and tin. Although these findings can be anticipated based on the large difference between the atomic radii of germanium and tin, the purpose of this work is to quantify the magnitude of the energies required to be overcome to stabilise thin films. By coherent epitaxial growth, the lattice spacing parallel to a substrate can to a large extent be constrained within a thin film and result in stabilisation.\cite{Wood1988,Zunger1989,Ozolins1998b} The extent to which coherent growth can stabilise germanium tin alloys is examined by constraining lattice parameters for the SQS along a selected plane to mimic the effect of growth on substrates for which epitaxially stable films of $\alpha$-Sn and/or germanium-tin alloys have already been demonstrated, namely: $\mathrm{Ge}$ (0.564 nm), $\mathrm{ZnTe/GaSb}$ (0.610 nm), and $\mathrm{CdTe/InSb}$ (0.648 nm).~\cite{Farrow1981,Hoffman1989,John1989,Bowman1990,1990PS,Piao1990} To explore the most promising orientations for stabilising the alloys by coherent growth, an epitaxial softening function is defined as
\begin{equation} \label{eqn:epi_Q}
q(a_S, \hat{G} ) = \frac{\Delta E^{epi}(a_{S}, \hat{G})}{\Delta E^{bulk}(a_{S})},
\end{equation}
which is the ratio between the increase in energy due to biaxial strain as a result of epitaxial growth along a direction $\hat{G}$ on a substrate with lattice constant $a_S$, and the increase in energy due to hydrostatically straining to the substrate lattice constant $a_S$. The ratio between the two energies quantifies the degree to which out-of-plane relaxation can stabilise an epitaxially grown thin film, with lower values of  $q(a_S,\hat{G})$ signalling increased stability. Lower values of the softening function in general imply a reduction in the formation of dislocations and other strain-induced film or surface defects.\cite{Ozolins1998a} \Cref{fig:Epi_soft_funcs} shows the epitaxial softening functions of germanium and tin as computed with DFT; it is seen that $\langle 100 \rangle$ is the \emph{softest} direction for germanium and tin for values of $a_S$ in the vicinity of their equilibrium lattice constants. This conclusion is retained at all compositions if the alloy epitaxial softening function is estimated as a weighted sum of the elemental softening functions. This suggests that $\langle 100 \rangle$-oriented films will tend to minimise internal strain energy and thus structural defects for epitaxially coherent $\mathrm{Ge_{1-x}Sn_{x}}$ films, and is consistent with at least one experimental analysis.~\cite{Asano2014} \\

\begin{figure}
\includegraphics[width=\columnwidth]{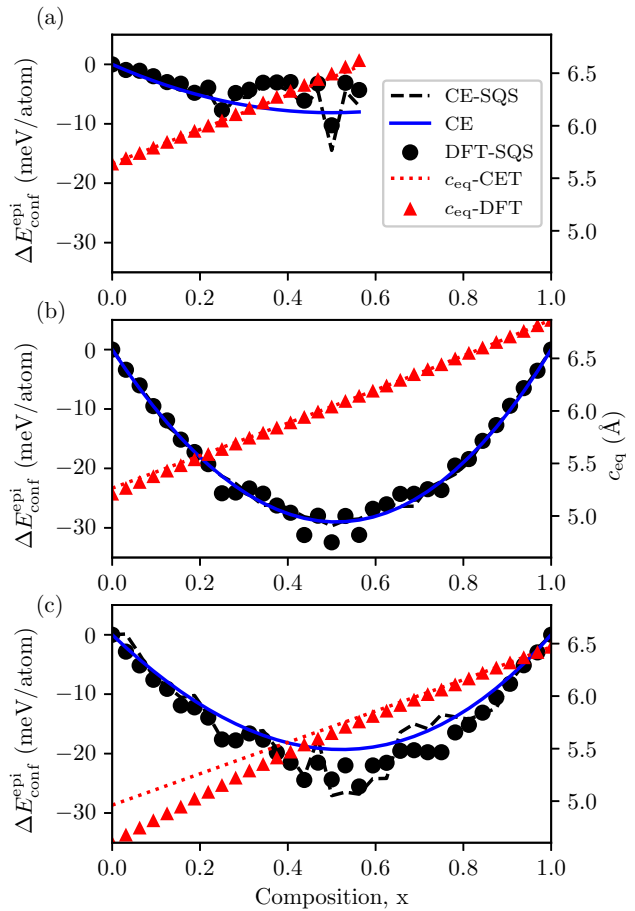} 
\centering
\caption{For alloys grown on (a) Ge, (b) ZnTe/GaSb, and (c) CdTe/InSb substrates: out of plane lattice constant $c_{\mathrm{eq}}$ calculated from DFT (red triangles) and continuum elasticity theory (red dots) referenced to the right axis, and the formation energy with respect to coherent decomposition to elemental constituents as calculated from the CE for random alloys (blue line) and using 64-atom SQS (black dashed line), and directly from DFT calculations for SQS (black dots). All formation energies are referenced to the left axis.}
\label{fig:Epi_CE}
\end{figure}

Epitaxial growth on (001)-oriented substrates is evaluated by constraining the lattice parameters of 64-atom SQS along $[100]$ and $[010]$ to match the lattice spacing to that of the three substrates under consideration. The third lattice parameter for each supercell is allowed to relax. This scheme neglects the role of surface and interfacial effects, but it does provides a baseline for the energetics of the alloy compositions independent of these effects.  \Cref{fig:Epi_CE} shows the computed epitaxial formation energy relative to decomposition into the coherently strained elemental constituents. In this case, the substrates suppress the tendency to segregate due to the coherent strain assumed for the segregated elemental components leading to a stabilisation of the alloys even at zero temperature. Note geometry optimizations for SQS biaxially strained to the Ge$(001)$ lattice constant with compositions above $\mathrm{x=0.56}$ resulted in amorphous structures and thus are not included in the analysis of \Cref{fig:Epi_CE}(a). Formation energies obtained from DFT calculations have been used to fit cluster expansion ECIs in order to investigate the case of random alloys and eliminate restrictions on the cluster ensemble averages arising from the use of the SQS approach. The CE predict all alloys to be energetically favourable with respect to coherent decomposition with a trend whereby stability decreases along the (ZnTe/GaSb)-(CdTe/InSb)-Ge sequence, indicating growth on germanium to be the least favourable choice for alloy growth if coherent decomposition is assumed. Additionally, \cref{fig:Epi_CE} includes the magnitude of the out-of-plane lattice parameter $c_{\mathrm{eq}}$ for each substrate as computed with DFT for SQS and a continuum elasticity theory (CET),~\cite{Zunger1989} where good agreement between both methods is found across most of the composition range in Ge and ZnTe/GaSb substrates. However, significant deviations between the two approaches are seen at low tin content alloys coherent to CdTe/InSb substrates due to anharmonic effects at these large strains not taken into account within CET. From analysis of the vibrational contributions to the free energy in a manner similar to \cref{fig:f_vs_t}, it is found that the aforementioned sequence in pseudomorphic alloy stability is maintained at finite temperatures, although we note that the bond stiffness vs. bond length model and harmonic approximation employed in this work are inadequate to describe the vibrational properties of pseudomorphic germanium and tin with up to 15\% biaxial strain, as evident from $c_{\mathrm{eq}}$ calculations employing CET.\\

Decompositions of epitaxial formation energies akin to those defined for bulk alloys in \cref{eq:decomp} are presented in \cref{tbl:epi_vol_decomp} for alloys at a composition $\mathrm{x=0.5}$. The effects of growth on lattice-matched substrates can be observed in the formation energy decomposition corresponding to the ZnTe/GaSb substrates. Whereas the computed epitaxial spin-flip $\delta E^{\mathrm{chem}, \mathrm{epi}}$ and internal relaxation $\delta E^{\mathrm{int,epi}}$ energies remain similar to the values found for bulk alloys, the volume deformation energy relative to epitaxially coherent decomposition $\Delta E_{VD}^{\mathrm{epi}}$ is reduced by $37\%$ resulting in the observed enhancement to stability. By comparing results across the different substrates, it is observed that destabilisation of component segregation results in an overall reduction of $\Delta E_{VD}^{\mathrm{epi}}$ with increasing substrate lattice parameter due to germanium's larger bulk modulus. However, the magnitudes of stabilising contributions $\delta E_{UR}^{\mathrm{chem}, \mathrm{epi}}$ and $\delta E^{\mathrm{int,epi}}$ are observed to also decrease with increasing bond lengths associated with alloys grown on substrates with larger lattice spacings. This leads to the ZnTe/GaSb substrate exhibiting the lowest value of $\Delta E_{\mathrm{conf}}^{\mathrm{epi}}$ across the three substrates. \\

\begin{figure}%[t]
\includegraphics[width=\columnwidth]{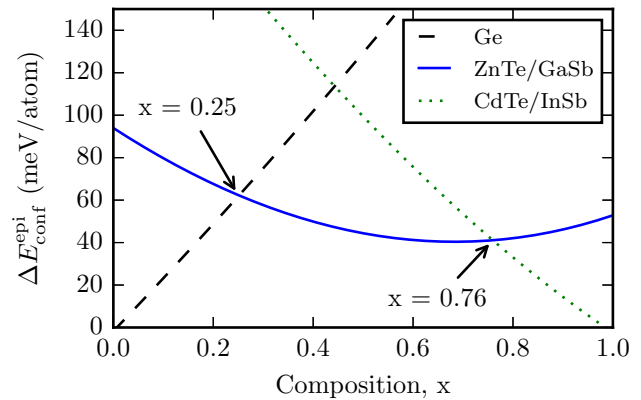} 
%\centering
\caption{Formation energy of alloys grown on Ge, ZnTe/GaSb, and CdTe/InSb substrates with respect to non-coherent decomposition into bulk components.}
\label{fig:Epi_bulk}
\end{figure}

As a final point, the relative stability of the alloys grown on different substrates in \cref{fig:Epi_bulk} are compared by computing the formation energy for epitaxial alloys with respect to non-coherent decomposition into their (unstrained) bulk elemental components. The substrate providing the lowest formation energy, and thus highest stability with respect to segregation, is observed to depend on alloy composition. Within the set of substrates considered, Ge is preferred for compositions below $\mathrm{x=0.25}$, CdTe/InSb for compositions above $\mathrm{x=0.76}$, and ZnTe/GaSb for $0.25 < \mathrm{x} < 0.76 $, indicating the latter to be energetically favourable across much of the composition range where the alloy exhibits are expected to display semimetallic behaviour. The asymmetry discussed for bulk alloys whereby tin-rich compositions have relatively lower formation energies is also present for alloys grown on ZnTe/GaSb, as seen in \cref{fig:Epi_bulk}.

\begin{table}[t]
\begin{ruledtabular}
\begin{tabular}{ccccc}

Substrate & $\Delta E_{\mathrm{conf}}^{\mathrm{epi}}$ & $\Delta E^{\mathrm{epi}}_{VD} $ & $\delta E^{chem,\mathrm{epi}}_{UR}$ & $\delta E^{int,\mathrm{epi}}$ \\
& \multicolumn{4}{c}{(meV/atom)} \\
	\hline
Ge & -10 & 195 & -107 & -97 \\
ZnTe/GaSb & -32 & 131 & -89 & -75 \\
CdTe/InSb & -24 & 114 & -76 & -62 \\

\end{tabular}
\end{ruledtabular}
 \caption{Epitaxial formation energy decomposition for alloys with composition $\mathrm{x=0.5}$ grown on each of the substrates included in this study.}
\label{tbl:epi_vol_decomp}
\end{table}	

\section{\label{Conclusion}Conclusion}
A study for the thermodynamic stability of GeSn alloys has been undertaken with models for both relaxed and pseudomorphic thick GeSn layers. A CE fit is performed and demonstrated to be capable of reproducing energies from DFT calculations for SQS across the full alloy composition range. From the CE, an asymmetry in the formation energies of random alloys as a function of alloy composition is identified whereby introduction of Sn into Ge-rich alloys results in larger formation energies relative to the case of introducing Ge into Sn-rich alloys. A partitioning of the formation energy into a volume deformation, chemical and relaxation terms reveals the volume deformation energy dominates the zero-temperature formation energies and drives immiscibility for these alloys with the chemical and relaxation energies serving to reduce the formation energies. Evaluation of free energy at finite temperatures reveals the configurational entropy dominates stabilising contributions with vibrational terms contributing to a lesser extent, and the electronic contributions have negligible impact in the temperature range of interest. CE fits describing configurational and temperature-dependent vibrational properties are employed to calculate the system's phase diagram using lattice Monte Carlo simulations. It is shown that the critical temperature for stability of Ge$_\mathrm{1-x}$Sn$_\mathrm{x}$ alloys generally decreases at higher Sn composition as a result of the aforementioned asymmetry in the formation energy versus composition. As well, vibrational terms introduce an opposing asymmetry in the phase diagram whereby the stability of Ge-rich alloys is enhanced at finite temperatures due to the softening of phonon modes associated with incorporation of Sn into the lattice. The phase diagram identifies critical temperatures between the range of the melting temperatures for Sn and Ge for essentially the entire composition range, in good agreement with known experimental properties for bulk GeSn alloys. It should be noted that relaxation in thin films or nanowires can be enhanced by strain relaxation normal to the surfaces and can serve to stabilise nanostructures. \\

The influence of pseudomorphic growth on stability was considered by studying alloys biaxially strained to the lattice spacing of substrates previously used for coherent epitaxial growth of Ge$_\mathrm{1-x}$Sn$_\mathrm{x}$ which provide lattice matching at compositions x=0, and approximately for x=0.5 and x=1. The preferred growth orientation is analysed with the epitaxial softening function which indicates the (001)-orientation is preferable at small to moderate strains. Analysis of the formation energies reveals pseudomorphic growth enhances alloy stability by impeding coherent segregation resulting in alloy stability even at zero temperature. Growth on ZnTe/GaSb substrates is predicted to provide the greatest epitaxial stabilisation across most of the composition range. Comparing pseudomorphic formation energies across substrates with respect to incoherent segregation reveals ZnTe/GaSb to be energetically favourable across approximately half of the composition range  ($\mathrm{0.25<x<0.76}$), with Ge (CdTe/InSb) substrates providing greater stability for $\mathrm{x<0.25}$ ($\mathrm{x>0.76}$). This result correlates with a recent study where similar crossing points were determined when computing thin film critical thicknesses using continuum elasticity theory.~\cite{BandGapPaper} It is worth noting that even though, for example, the results for the formation energies for an x=0.25 alloy grown on either Ge or ZnTe/GaSb substrates are predicted to be comparable, their resulting electronic properties can be significantly different. Recent calculations report the compressive strain built into Ge$_{0.75}$Sn$_{0.25}$/Ge results in a zero band gap semimetal band structure, whereas the tensile strain built into Ge$_{0.75}$Sn$_{0.25}$/ZnTe leads to a band overlap (\emph{negative} band gap) of approximately 600 meV.~\cite{BandGapPaper} \\

Results presented in this work emphasise built-in strain to be the driver of immiscibility in Ge$_\mathrm{1-x}$Sn$_\mathrm{x}$ alloys. Strategies for stabilisation must thus be directed at tackling this issue by balancing several complementary effects such as relaxation normal to surfaces in nanostructures, by targeting Sn-rich compositions, and by choice of substrates that induce epitaxial stabilisation of alloys while delivering suitable critical thicknesses and electronic properties. In the context of materials engineering, selection of alloy composition, strain, film thickness (quantum confinement), and surface chemistry enables a wide range of options for the tailoring of the thermomechanical and electronic properties of GeSn alloys for technological applications. 

\begin{acknowledgments}
This work was funded by Science Foundation Ireland through a Principal Investigator award Grant No. 13/IA/1956, by the Nottingham Ningbo New Materials Institute, and the National Science Foundation of China Project Code 61974079.
\end{acknowledgments}


\begin{thebibliography}{78}%
\makeatletter
\providecommand \@ifxundefined [1]{%
 \@ifx{#1\undefined}
}%
\providecommand \@ifnum [1]{%
 \ifnum #1\expandafter \@firstoftwo
 \else \expandafter \@secondoftwo
 \fi
}%
\providecommand \@ifx [1]{%
 \ifx #1\expandafter \@firstoftwo
 \else \expandafter \@secondoftwo
 \fi
}%
\providecommand \natexlab [1]{#1}%
\providecommand \enquote  [1]{``#1''}%
\providecommand \bibnamefont  [1]{#1}%
\providecommand \bibfnamefont [1]{#1}%
\providecommand \citenamefont [1]{#1}%
\providecommand \href@noop [0]{\@secondoftwo}%
\providecommand \href [0]{\begingroup \@sanitize@url \@href}%
\providecommand \@href[1]{\@@startlink{#1}\@@href}%
\providecommand \@@href[1]{\endgroup#1\@@endlink}%
\providecommand \@sanitize@url [0]{\catcode `\\12\catcode `\$12\catcode
  `\&12\catcode `\#12\catcode `\^12\catcode `\_12\catcode `\%12\relax}%
\providecommand \@@startlink[1]{}%
\providecommand \@@endlink[0]{}%
\providecommand \url  [0]{\begingroup\@sanitize@url \@url }%
\providecommand \@url [1]{\endgroup\@href {#1}{\urlprefix }}%
\providecommand \urlprefix  [0]{URL }%
\providecommand \Eprint [0]{\href }%
\providecommand \doibase [0]{http://dx.doi.org/}%
\providecommand \selectlanguage [0]{\@gobble}%
\providecommand \bibinfo  [0]{\@secondoftwo}%
\providecommand \bibfield  [0]{\@secondoftwo}%
\providecommand \translation [1]{[#1]}%
\providecommand \BibitemOpen [0]{}%
\providecommand \bibitemStop [0]{}%
\providecommand \bibitemNoStop [0]{.\EOS\space}%
\providecommand \EOS [0]{\spacefactor3000\relax}%
\providecommand \BibitemShut  [1]{\csname bibitem#1\endcsname}%
\let\auto@bib@innerbib\@empty
%</preamble>
\bibitem [{\citenamefont {Doherty}\ \emph {et~al.}(2020)\citenamefont
  {Doherty}, \citenamefont {Biswas}, \citenamefont {Galluccio}, \citenamefont
  {Broderick}, \citenamefont {Garcia-Gil}, \citenamefont {Duffy}, \citenamefont
  {O'Reilly},\ and\ \citenamefont {Holmes}}]{Doherty2020}%
  \BibitemOpen
  \bibfield  {author} {\bibinfo {author} {\bibfnamefont {J.}~\bibnamefont
  {Doherty}}, \bibinfo {author} {\bibfnamefont {S.}~\bibnamefont {Biswas}},
  \bibinfo {author} {\bibfnamefont {E.}~\bibnamefont {Galluccio}}, \bibinfo
  {author} {\bibfnamefont {C.~A.}\ \bibnamefont {Broderick}}, \bibinfo {author}
  {\bibfnamefont {A.}~\bibnamefont {Garcia-Gil}}, \bibinfo {author}
  {\bibfnamefont {R.}~\bibnamefont {Duffy}}, \bibinfo {author} {\bibfnamefont
  {E.~P.}\ \bibnamefont {O'Reilly}}, \ and\ \bibinfo {author} {\bibfnamefont
  {J.~D.}\ \bibnamefont {Holmes}},\ }\href {\doibase
  10.1021/acs.chemmater.9b04136} {\bibfield  {journal} {\bibinfo  {journal}
  {Chemistry of Materials}\ }\textbf {\bibinfo {volume} {32}},\ \bibinfo
  {pages} {4383} (\bibinfo {year} {2020})}\BibitemShut {NoStop}%
\bibitem [{\citenamefont {Wang}\ \emph {et~al.}(2007)\citenamefont {Wang},
  \citenamefont {Toh}, \citenamefont {Wang}, \citenamefont {Seng},
  \citenamefont {Tripathy}, \citenamefont {Osipowicz}, \citenamefont {Chan},
  \citenamefont {Hoe}, \citenamefont {Balakumar}, \citenamefont {Tung},
  \citenamefont {Lo}, \citenamefont {Samudra},\ and\ \citenamefont
  {Yeo}}]{Wang2007}%
  \BibitemOpen
  \bibfield  {author} {\bibinfo {author} {\bibfnamefont {G.~H.}\ \bibnamefont
  {Wang}}, \bibinfo {author} {\bibfnamefont {E.-H.}\ \bibnamefont {Toh}},
  \bibinfo {author} {\bibfnamefont {X.}~\bibnamefont {Wang}}, \bibinfo {author}
  {\bibfnamefont {D.~H.~L.}\ \bibnamefont {Seng}}, \bibinfo {author}
  {\bibfnamefont {S.}~\bibnamefont {Tripathy}}, \bibinfo {author}
  {\bibfnamefont {T.}~\bibnamefont {Osipowicz}}, \bibinfo {author}
  {\bibfnamefont {T.~K.}\ \bibnamefont {Chan}}, \bibinfo {author}
  {\bibfnamefont {K.~M.}\ \bibnamefont {Hoe}}, \bibinfo {author} {\bibfnamefont
  {S.}~\bibnamefont {Balakumar}}, \bibinfo {author} {\bibfnamefont {C.~H.}\
  \bibnamefont {Tung}}, \bibinfo {author} {\bibfnamefont {G.-Q.}\ \bibnamefont
  {Lo}}, \bibinfo {author} {\bibfnamefont {G.}~\bibnamefont {Samudra}}, \ and\
  \bibinfo {author} {\bibfnamefont {Y.-C.}\ \bibnamefont {Yeo}},\ }in\ \href
  {\doibase 10.1109/iedm.2007.4418882} {\emph {\bibinfo {booktitle} {2007
  {IEEE} International Electron Devices Meeting}}}\ (\bibinfo  {publisher}
  {{IEEE}},\ \bibinfo {year} {2007})\BibitemShut {NoStop}%
\bibitem [{\citenamefont {Conley}\ \emph {et~al.}(2014)\citenamefont {Conley},
  \citenamefont {Mosleh}, \citenamefont {Ghetmiri}, \citenamefont {Du},
  \citenamefont {Soref}, \citenamefont {Sun}, \citenamefont {Margetis},
  \citenamefont {Tolle}, \citenamefont {Naseem},\ and\ \citenamefont
  {Yu}}]{Conley2014}%
  \BibitemOpen
  \bibfield  {author} {\bibinfo {author} {\bibfnamefont {B.~R.}\ \bibnamefont
  {Conley}}, \bibinfo {author} {\bibfnamefont {A.}~\bibnamefont {Mosleh}},
  \bibinfo {author} {\bibfnamefont {S.~A.}\ \bibnamefont {Ghetmiri}}, \bibinfo
  {author} {\bibfnamefont {W.}~\bibnamefont {Du}}, \bibinfo {author}
  {\bibfnamefont {R.~A.}\ \bibnamefont {Soref}}, \bibinfo {author}
  {\bibfnamefont {G.}~\bibnamefont {Sun}}, \bibinfo {author} {\bibfnamefont
  {J.}~\bibnamefont {Margetis}}, \bibinfo {author} {\bibfnamefont
  {J.}~\bibnamefont {Tolle}}, \bibinfo {author} {\bibfnamefont {H.~A.}\
  \bibnamefont {Naseem}}, \ and\ \bibinfo {author} {\bibfnamefont {S.-Q.}\
  \bibnamefont {Yu}},\ }\href {\doibase 10.1364/oe.22.015639} {\bibfield
  {journal} {\bibinfo  {journal} {Optics Express}\ }\textbf {\bibinfo {volume}
  {22}},\ \bibinfo {pages} {15639} (\bibinfo {year} {2014})}\BibitemShut
  {NoStop}%
\bibitem [{\citenamefont {Olesinski}\ and\ \citenamefont
  {Abbaschian}(1984)}]{OlesinskiAbbaschian1984}%
  \BibitemOpen
  \bibfield  {author} {\bibinfo {author} {\bibfnamefont {R.~W.}\ \bibnamefont
  {Olesinski}}\ and\ \bibinfo {author} {\bibfnamefont {G.~J.}\ \bibnamefont
  {Abbaschian}},\ }\href {\doibase 10.1007/bf02868550} {\bibfield  {journal}
  {\bibinfo  {journal} {Bulletin of Alloy Phase Diagrams}\ }\textbf {\bibinfo
  {volume} {5}},\ \bibinfo {pages} {265} (\bibinfo {year} {1984})}\BibitemShut
  {NoStop}%
\bibitem [{\citenamefont {Raynor}\ and\ \citenamefont
  {Smith}(1958)}]{Raynor1958a}%
  \BibitemOpen
  \bibfield  {author} {\bibinfo {author} {\bibfnamefont {G.~V.}\ \bibnamefont
  {Raynor}}\ and\ \bibinfo {author} {\bibfnamefont {R.~W.}\ \bibnamefont
  {Smith}},\ }\href {\doibase 10.1098/rspa.1958.0028} {\bibfield  {journal}
  {\bibinfo  {journal} {Proceedings of the Royal Society of London. Series A.
  Mathematical and Physical Sciences}\ }\textbf {\bibinfo {volume} {244}},\
  \bibinfo {pages} {101} (\bibinfo {year} {1958})}\BibitemShut {NoStop}%
\bibitem [{\citenamefont {Piao}\ \emph {et~al.}(1990)\citenamefont {Piao},
  \citenamefont {Beresford}, \citenamefont {Licata}, \citenamefont {Wang},\
  and\ \citenamefont {Homma}}]{Piao1990}%
  \BibitemOpen
  \bibfield  {author} {\bibinfo {author} {\bibfnamefont {J.}~\bibnamefont
  {Piao}}, \bibinfo {author} {\bibfnamefont {R.}~\bibnamefont {Beresford}},
  \bibinfo {author} {\bibfnamefont {T.}~\bibnamefont {Licata}}, \bibinfo
  {author} {\bibfnamefont {W.}~\bibnamefont {Wang}}, \ and\ \bibinfo {author}
  {\bibfnamefont {H.}~\bibnamefont {Homma}},\ }\href {\doibase
  doi.org/10.1116/1.584814} {\bibfield  {journal} {\bibinfo  {journal} {Journal
  of Vacuum Science \& Technology B: Microelectronics Processing and
  Phenomena}\ }\textbf {\bibinfo {volume} {8}},\ \bibinfo {pages} {221}
  (\bibinfo {year} {1990})}\BibitemShut {NoStop}%
\bibitem [{\citenamefont {Gencarelli}\ \emph {et~al.}(2013)\citenamefont
  {Gencarelli}, \citenamefont {Vincent}, \citenamefont {Demeulemeester},
  \citenamefont {Vantomme}, \citenamefont {Moussa}, \citenamefont {Franquet},
  \citenamefont {Kumar}, \citenamefont {Bender}, \citenamefont {Meersschaut},
  \citenamefont {Vandervorst}, \citenamefont {Loo}, \citenamefont {Caymax},
  \citenamefont {Temst},\ and\ \citenamefont {Heyns}}]{Gencarelli2013}%
  \BibitemOpen
  \bibfield  {author} {\bibinfo {author} {\bibfnamefont {F.}~\bibnamefont
  {Gencarelli}}, \bibinfo {author} {\bibfnamefont {B.}~\bibnamefont {Vincent}},
  \bibinfo {author} {\bibfnamefont {J.}~\bibnamefont {Demeulemeester}},
  \bibinfo {author} {\bibfnamefont {A.}~\bibnamefont {Vantomme}}, \bibinfo
  {author} {\bibfnamefont {A.}~\bibnamefont {Moussa}}, \bibinfo {author}
  {\bibfnamefont {A.}~\bibnamefont {Franquet}}, \bibinfo {author}
  {\bibfnamefont {A.}~\bibnamefont {Kumar}}, \bibinfo {author} {\bibfnamefont
  {H.}~\bibnamefont {Bender}}, \bibinfo {author} {\bibfnamefont
  {J.}~\bibnamefont {Meersschaut}}, \bibinfo {author} {\bibfnamefont
  {W.}~\bibnamefont {Vandervorst}}, \bibinfo {author} {\bibfnamefont
  {R.}~\bibnamefont {Loo}}, \bibinfo {author} {\bibfnamefont {M.}~\bibnamefont
  {Caymax}}, \bibinfo {author} {\bibfnamefont {K.}~\bibnamefont {Temst}}, \
  and\ \bibinfo {author} {\bibfnamefont {M.}~\bibnamefont {Heyns}},\ }\href
  {\doibase 10.1149/05009.0875ecst} {\bibfield  {journal} {\bibinfo  {journal}
  {{ECS} Transactions}\ }\textbf {\bibinfo {volume} {50}},\ \bibinfo {pages}
  {875} (\bibinfo {year} {2013})}\BibitemShut {NoStop}%
\bibitem [{\citenamefont {Wirths}\ \emph {et~al.}(2016)\citenamefont {Wirths},
  \citenamefont {Buca},\ and\ \citenamefont {Mantl}}]{WirthsBucaMantl2016}%
  \BibitemOpen
  \bibfield  {author} {\bibinfo {author} {\bibfnamefont {S.}~\bibnamefont
  {Wirths}}, \bibinfo {author} {\bibfnamefont {D.}~\bibnamefont {Buca}}, \ and\
  \bibinfo {author} {\bibfnamefont {S.}~\bibnamefont {Mantl}},\ }\href
  {\doibase 10.1016/j.pcrysgrow.2015.11.001} {\bibfield  {journal} {\bibinfo
  {journal} {Progress in Crystal Growth and Characterization of Materials}\
  }\textbf {\bibinfo {volume} {62}},\ \bibinfo {pages} {1} (\bibinfo {year}
  {2016})}\BibitemShut {NoStop}%
\bibitem [{\citenamefont {Dou}\ \emph {et~al.}(2018)\citenamefont {Dou},
  \citenamefont {Benamara}, \citenamefont {Mosleh}, \citenamefont {Margetis},
  \citenamefont {Grant}, \citenamefont {Zhou}, \citenamefont {Al-Kabi},
  \citenamefont {Du}, \citenamefont {Tolle}, \citenamefont {Li} \emph
  {et~al.}}]{Dou2018}%
  \BibitemOpen
  \bibfield  {author} {\bibinfo {author} {\bibfnamefont {W.}~\bibnamefont
  {Dou}}, \bibinfo {author} {\bibfnamefont {M.}~\bibnamefont {Benamara}},
  \bibinfo {author} {\bibfnamefont {A.}~\bibnamefont {Mosleh}}, \bibinfo
  {author} {\bibfnamefont {J.}~\bibnamefont {Margetis}}, \bibinfo {author}
  {\bibfnamefont {P.}~\bibnamefont {Grant}}, \bibinfo {author} {\bibfnamefont
  {Y.}~\bibnamefont {Zhou}}, \bibinfo {author} {\bibfnamefont {S.}~\bibnamefont
  {Al-Kabi}}, \bibinfo {author} {\bibfnamefont {W.}~\bibnamefont {Du}},
  \bibinfo {author} {\bibfnamefont {J.}~\bibnamefont {Tolle}}, \bibinfo
  {author} {\bibfnamefont {B.}~\bibnamefont {Li}},  \emph {et~al.},\ }\href
  {\doibase doi.org/10.1038/s41598-018-24018-6} {\bibfield  {journal} {\bibinfo
   {journal} {Scientific reports}\ }\textbf {\bibinfo {volume} {8}},\ \bibinfo
  {pages} {5640} (\bibinfo {year} {2018})}\BibitemShut {NoStop}%
\bibitem [{\citenamefont {Wei}\ \emph {et~al.}(2021)\citenamefont {Wei},
  \citenamefont {Miao}, \citenamefont {Pan}, \citenamefont {wei Zhang},
  \citenamefont {Li}, \citenamefont {Lu},\ and\ \citenamefont
  {Chen}}]{WEI2021125996}%
  \BibitemOpen
  \bibfield  {author} {\bibinfo {author} {\bibfnamefont {L.}~\bibnamefont
  {Wei}}, \bibinfo {author} {\bibfnamefont {Y.}~\bibnamefont {Miao}}, \bibinfo
  {author} {\bibfnamefont {R.}~\bibnamefont {Pan}}, \bibinfo {author}
  {\bibfnamefont {W.}~\bibnamefont {wei Zhang}}, \bibinfo {author}
  {\bibfnamefont {C.}~\bibnamefont {Li}}, \bibinfo {author} {\bibfnamefont
  {H.}~\bibnamefont {Lu}}, \ and\ \bibinfo {author} {\bibfnamefont {Y.-F.}\
  \bibnamefont {Chen}},\ }\href {\doibase 10.1016/j.jcrysgro.2020.125996}
  {\bibfield  {journal} {\bibinfo  {journal} {Journal of Crystal Growth}\
  }\textbf {\bibinfo {volume} {557}},\ \bibinfo {pages} {125996} (\bibinfo
  {year} {2021})}\BibitemShut {NoStop}%
\bibitem [{\citenamefont {Dascalescu}\ \emph {et~al.}(2020)\citenamefont
  {Dascalescu}, \citenamefont {Zoita}, \citenamefont {Slav}, \citenamefont
  {Matei}, \citenamefont {Iftimie}, \citenamefont {Comanescu}, \citenamefont
  {Lepadatu}, \citenamefont {Palade}, \citenamefont {Lazanu}, \citenamefont
  {Buca}, \citenamefont {Teodorescu}, \citenamefont {Ciurea}, \citenamefont
  {Braic},\ and\ \citenamefont {Stoica}}]{Dascalescu2020}%
  \BibitemOpen
  \bibfield  {author} {\bibinfo {author} {\bibfnamefont {I.}~\bibnamefont
  {Dascalescu}}, \bibinfo {author} {\bibfnamefont {N.~C.}\ \bibnamefont
  {Zoita}}, \bibinfo {author} {\bibfnamefont {A.}~\bibnamefont {Slav}},
  \bibinfo {author} {\bibfnamefont {E.}~\bibnamefont {Matei}}, \bibinfo
  {author} {\bibfnamefont {S.}~\bibnamefont {Iftimie}}, \bibinfo {author}
  {\bibfnamefont {F.}~\bibnamefont {Comanescu}}, \bibinfo {author}
  {\bibfnamefont {A.-M.}\ \bibnamefont {Lepadatu}}, \bibinfo {author}
  {\bibfnamefont {C.}~\bibnamefont {Palade}}, \bibinfo {author} {\bibfnamefont
  {S.}~\bibnamefont {Lazanu}}, \bibinfo {author} {\bibfnamefont
  {D.}~\bibnamefont {Buca}}, \bibinfo {author} {\bibfnamefont {V.~S.}\
  \bibnamefont {Teodorescu}}, \bibinfo {author} {\bibfnamefont {M.~L.}\
  \bibnamefont {Ciurea}}, \bibinfo {author} {\bibfnamefont {M.}~\bibnamefont
  {Braic}}, \ and\ \bibinfo {author} {\bibfnamefont {T.}~\bibnamefont
  {Stoica}},\ }\href {\doibase 10.1021/acsami.0c06212} {\bibfield  {journal}
  {\bibinfo  {journal} {{ACS} Applied Materials {\&} Interfaces}\ }\textbf
  {\bibinfo {volume} {12}},\ \bibinfo {pages} {33879} (\bibinfo {year}
  {2020})}\BibitemShut {NoStop}%
\bibitem [{\citenamefont {Tai}\ \emph {et~al.}(2020)\citenamefont {Tai},
  \citenamefont {Yeh}, \citenamefont {An}, \citenamefont {Cheng}, \citenamefont
  {Kim},\ and\ \citenamefont {Chang}}]{Tai2020}%
  \BibitemOpen
  \bibfield  {author} {\bibinfo {author} {\bibfnamefont {Y.-C.}\ \bibnamefont
  {Tai}}, \bibinfo {author} {\bibfnamefont {P.-L.}\ \bibnamefont {Yeh}},
  \bibinfo {author} {\bibfnamefont {S.}~\bibnamefont {An}}, \bibinfo {author}
  {\bibfnamefont {H.-H.}\ \bibnamefont {Cheng}}, \bibinfo {author}
  {\bibfnamefont {M.}~\bibnamefont {Kim}}, \ and\ \bibinfo {author}
  {\bibfnamefont {G.-E.}\ \bibnamefont {Chang}},\ }\href {\doibase
  10.1088/1361-6528/aba6b1} {\bibfield  {journal} {\bibinfo  {journal}
  {Nanotechnology}\ }\textbf {\bibinfo {volume} {31}},\ \bibinfo {pages}
  {445301} (\bibinfo {year} {2020})}\BibitemShut {NoStop}%
\bibitem [{\citenamefont {Wirths}\ \emph {et~al.}(2015)\citenamefont {Wirths},
  \citenamefont {Geiger}, \citenamefont {von~den Driesch}, \citenamefont
  {Mussler}, \citenamefont {Stoica}, \citenamefont {Mantl}, \citenamefont
  {Ikonic}, \citenamefont {Luysberg}, \citenamefont {Chiussi}, \citenamefont
  {Hartmann}, \citenamefont {Sigg}, \citenamefont {Faist}, \citenamefont
  {Buca},\ and\ \citenamefont {Grutzmacher}}]{Wirths2015}%
  \BibitemOpen
  \bibfield  {author} {\bibinfo {author} {\bibfnamefont {S.}~\bibnamefont
  {Wirths}}, \bibinfo {author} {\bibfnamefont {R.}~\bibnamefont {Geiger}},
  \bibinfo {author} {\bibfnamefont {N.}~\bibnamefont {von~den Driesch}},
  \bibinfo {author} {\bibfnamefont {G.}~\bibnamefont {Mussler}}, \bibinfo
  {author} {\bibfnamefont {T.}~\bibnamefont {Stoica}}, \bibinfo {author}
  {\bibfnamefont {S.}~\bibnamefont {Mantl}}, \bibinfo {author} {\bibfnamefont
  {Z.}~\bibnamefont {Ikonic}}, \bibinfo {author} {\bibfnamefont
  {M.}~\bibnamefont {Luysberg}}, \bibinfo {author} {\bibfnamefont
  {S.}~\bibnamefont {Chiussi}}, \bibinfo {author} {\bibfnamefont {J.~M.}\
  \bibnamefont {Hartmann}}, \bibinfo {author} {\bibfnamefont {H.}~\bibnamefont
  {Sigg}}, \bibinfo {author} {\bibfnamefont {J.}~\bibnamefont {Faist}},
  \bibinfo {author} {\bibfnamefont {D.}~\bibnamefont {Buca}}, \ and\ \bibinfo
  {author} {\bibfnamefont {D.}~\bibnamefont {Grutzmacher}},\ }\href {\doibase
  10.1038/nphoton.2014.321} {\bibfield  {journal} {\bibinfo  {journal} {Nature
  Photonics}\ }\textbf {\bibinfo {volume} {9}},\ \bibinfo {pages} {88}
  (\bibinfo {year} {2015})}\BibitemShut {NoStop}%
\bibitem [{\citenamefont {Reboud}\ \emph {et~al.}(2017)\citenamefont {Reboud},
  \citenamefont {Gassenq}, \citenamefont {Pauc}, \citenamefont {Aubin},
  \citenamefont {Milord}, \citenamefont {Thai}, \citenamefont {Bertrand},
  \citenamefont {Guilloy}, \citenamefont {Rouchon}, \citenamefont {Rothman}
  \emph {et~al.}}]{Reboud2017}%
  \BibitemOpen
  \bibfield  {author} {\bibinfo {author} {\bibfnamefont {V.}~\bibnamefont
  {Reboud}}, \bibinfo {author} {\bibfnamefont {A.}~\bibnamefont {Gassenq}},
  \bibinfo {author} {\bibfnamefont {N.}~\bibnamefont {Pauc}}, \bibinfo {author}
  {\bibfnamefont {J.}~\bibnamefont {Aubin}}, \bibinfo {author} {\bibfnamefont
  {L.}~\bibnamefont {Milord}}, \bibinfo {author} {\bibfnamefont
  {Q.}~\bibnamefont {Thai}}, \bibinfo {author} {\bibfnamefont {M.}~\bibnamefont
  {Bertrand}}, \bibinfo {author} {\bibfnamefont {K.}~\bibnamefont {Guilloy}},
  \bibinfo {author} {\bibfnamefont {D.}~\bibnamefont {Rouchon}}, \bibinfo
  {author} {\bibfnamefont {J.}~\bibnamefont {Rothman}},  \emph {et~al.},\
  }\href {\doibase 10.1063/1.5000353} {\bibfield  {journal} {\bibinfo
  {journal} {Applied Physics Letters}\ }\textbf {\bibinfo {volume} {111}},\
  \bibinfo {pages} {092101} (\bibinfo {year} {2017})}\BibitemShut {NoStop}%
\bibitem [{\citenamefont {Rainko}\ \emph {et~al.}(2019)\citenamefont {Rainko},
  \citenamefont {Ikonic}, \citenamefont {Elbaz}, \citenamefont {von~den
  Driesch}, \citenamefont {Stange}, \citenamefont {Herth}, \citenamefont
  {Boucaud}, \citenamefont {Kurdi}, \citenamefont {Grützmacher},\ and\
  \citenamefont {Buca}}]{Rainko2019}%
  \BibitemOpen
  \bibfield  {author} {\bibinfo {author} {\bibfnamefont {D.}~\bibnamefont
  {Rainko}}, \bibinfo {author} {\bibfnamefont {Z.}~\bibnamefont {Ikonic}},
  \bibinfo {author} {\bibfnamefont {A.}~\bibnamefont {Elbaz}}, \bibinfo
  {author} {\bibfnamefont {N.}~\bibnamefont {von~den Driesch}}, \bibinfo
  {author} {\bibfnamefont {D.}~\bibnamefont {Stange}}, \bibinfo {author}
  {\bibfnamefont {E.}~\bibnamefont {Herth}}, \bibinfo {author} {\bibfnamefont
  {P.}~\bibnamefont {Boucaud}}, \bibinfo {author} {\bibfnamefont {M.~E.}\
  \bibnamefont {Kurdi}}, \bibinfo {author} {\bibfnamefont {D.}~\bibnamefont
  {Grützmacher}}, \ and\ \bibinfo {author} {\bibfnamefont {D.}~\bibnamefont
  {Buca}},\ }\href {\doibase 10.1038/s41598-018-36837-8} {\bibfield  {journal}
  {\bibinfo  {journal} {Scientific Reports}\ }\textbf {\bibinfo {volume} {9}},\
  \bibinfo {pages} {259} (\bibinfo {year} {2019})}\BibitemShut {NoStop}%
\bibitem [{\citenamefont {Elbaz}\ \emph {et~al.}(2020)\citenamefont {Elbaz},
  \citenamefont {Buca}, \citenamefont {von~den Driesch}, \citenamefont
  {Pantzas}, \citenamefont {Patriarche}, \citenamefont {Zerounian},
  \citenamefont {Herth}, \citenamefont {Checoury}, \citenamefont {Sauvage},
  \citenamefont {Sagnes}, \citenamefont {Foti}, \citenamefont {Ossikovski},
  \citenamefont {Hartmann}, \citenamefont {Boeuf}, \citenamefont {Ikonic},
  \citenamefont {Boucaud}, \citenamefont {Grützmacher},\ and\ \citenamefont
  {Kurdi}}]{Elbaz2020}%
  \BibitemOpen
  \bibfield  {author} {\bibinfo {author} {\bibfnamefont {A.}~\bibnamefont
  {Elbaz}}, \bibinfo {author} {\bibfnamefont {D.}~\bibnamefont {Buca}},
  \bibinfo {author} {\bibfnamefont {N.}~\bibnamefont {von~den Driesch}},
  \bibinfo {author} {\bibfnamefont {K.}~\bibnamefont {Pantzas}}, \bibinfo
  {author} {\bibfnamefont {G.}~\bibnamefont {Patriarche}}, \bibinfo {author}
  {\bibfnamefont {N.}~\bibnamefont {Zerounian}}, \bibinfo {author}
  {\bibfnamefont {E.}~\bibnamefont {Herth}}, \bibinfo {author} {\bibfnamefont
  {X.}~\bibnamefont {Checoury}}, \bibinfo {author} {\bibfnamefont
  {S.}~\bibnamefont {Sauvage}}, \bibinfo {author} {\bibfnamefont
  {I.}~\bibnamefont {Sagnes}}, \bibinfo {author} {\bibfnamefont
  {A.}~\bibnamefont {Foti}}, \bibinfo {author} {\bibfnamefont {R.}~\bibnamefont
  {Ossikovski}}, \bibinfo {author} {\bibfnamefont {J.-M.}\ \bibnamefont
  {Hartmann}}, \bibinfo {author} {\bibfnamefont {F.}~\bibnamefont {Boeuf}},
  \bibinfo {author} {\bibfnamefont {Z.}~\bibnamefont {Ikonic}}, \bibinfo
  {author} {\bibfnamefont {P.}~\bibnamefont {Boucaud}}, \bibinfo {author}
  {\bibfnamefont {D.}~\bibnamefont {Grützmacher}}, \ and\ \bibinfo {author}
  {\bibfnamefont {M.~E.}\ \bibnamefont {Kurdi}},\ }\href {\doibase
  10.1038/s41566-020-0601-5} {\bibfield  {journal} {\bibinfo  {journal} {Nature
  Photonics}\ }\textbf {\bibinfo {volume} {14}},\ \bibinfo {pages} {375}
  (\bibinfo {year} {2020})}\BibitemShut {NoStop}%
\bibitem [{\citenamefont {Chr{\'{e}}tien}\ \emph {et~al.}(2019)\citenamefont
  {Chr{\'{e}}tien}, \citenamefont {Pauc}, \citenamefont {Pilon}, \citenamefont
  {Bertrand}, \citenamefont {Thai}, \citenamefont {Casiez}, \citenamefont
  {Bernier}, \citenamefont {Dansas}, \citenamefont {Gergaud}, \citenamefont
  {Delamadeleine}, \citenamefont {Khazaka}, \citenamefont {Sigg}, \citenamefont
  {Faist}, \citenamefont {Chelnokov}, \citenamefont {Reboud}, \citenamefont
  {Hartmann},\ and\ \citenamefont {Calvo}}]{Chretien2019}%
  \BibitemOpen
  \bibfield  {author} {\bibinfo {author} {\bibfnamefont {J.}~\bibnamefont
  {Chr{\'{e}}tien}}, \bibinfo {author} {\bibfnamefont {N.}~\bibnamefont
  {Pauc}}, \bibinfo {author} {\bibfnamefont {F.~A.}\ \bibnamefont {Pilon}},
  \bibinfo {author} {\bibfnamefont {M.}~\bibnamefont {Bertrand}}, \bibinfo
  {author} {\bibfnamefont {Q.-M.}\ \bibnamefont {Thai}}, \bibinfo {author}
  {\bibfnamefont {L.}~\bibnamefont {Casiez}}, \bibinfo {author} {\bibfnamefont
  {N.}~\bibnamefont {Bernier}}, \bibinfo {author} {\bibfnamefont
  {H.}~\bibnamefont {Dansas}}, \bibinfo {author} {\bibfnamefont
  {P.}~\bibnamefont {Gergaud}}, \bibinfo {author} {\bibfnamefont
  {E.}~\bibnamefont {Delamadeleine}}, \bibinfo {author} {\bibfnamefont
  {R.}~\bibnamefont {Khazaka}}, \bibinfo {author} {\bibfnamefont
  {H.}~\bibnamefont {Sigg}}, \bibinfo {author} {\bibfnamefont {J.}~\bibnamefont
  {Faist}}, \bibinfo {author} {\bibfnamefont {A.}~\bibnamefont {Chelnokov}},
  \bibinfo {author} {\bibfnamefont {V.}~\bibnamefont {Reboud}}, \bibinfo
  {author} {\bibfnamefont {J.-M.}\ \bibnamefont {Hartmann}}, \ and\ \bibinfo
  {author} {\bibfnamefont {V.}~\bibnamefont {Calvo}},\ }\href {\doibase
  10.1021/acsphotonics.9b00712} {\bibfield  {journal} {\bibinfo  {journal}
  {{ACS} Photonics}\ }\textbf {\bibinfo {volume} {6}},\ \bibinfo {pages} {2462}
  (\bibinfo {year} {2019})}\BibitemShut {NoStop}%
\bibitem [{\citenamefont {Polak}\ \emph {et~al.}(2017)\citenamefont {Polak},
  \citenamefont {Scharoch},\ and\ \citenamefont {Kudrawiec}}]{Polak2017}%
  \BibitemOpen
  \bibfield  {author} {\bibinfo {author} {\bibfnamefont {M.~P.}\ \bibnamefont
  {Polak}}, \bibinfo {author} {\bibfnamefont {P.}~\bibnamefont {Scharoch}}, \
  and\ \bibinfo {author} {\bibfnamefont {R.}~\bibnamefont {Kudrawiec}},\ }\href
  {\doibase 10.1088/1361-6463/aa67bf} {\bibfield  {journal} {\bibinfo
  {journal} {Journal of Physics D: Applied Physics}\ }\textbf {\bibinfo
  {volume} {50}},\ \bibinfo {pages} {195103} (\bibinfo {year}
  {2017})}\BibitemShut {NoStop}%
\bibitem [{\citenamefont {O'Donnell}\ \emph {et~al.}()\citenamefont
  {O'Donnell}, \citenamefont {Sanchez-Soares}, \citenamefont {Broderick},\ and\
  \citenamefont {Greer}}]{ODonnell2020}%
  \BibitemOpen
  \bibfield  {author} {\bibinfo {author} {\bibfnamefont {C.}~\bibnamefont
  {O'Donnell}}, \bibinfo {author} {\bibfnamefont {A.}~\bibnamefont
  {Sanchez-Soares}}, \bibinfo {author} {\bibfnamefont {C.~A.}\ \bibnamefont
  {Broderick}}, \ and\ \bibinfo {author} {\bibfnamefont {J.~C.}\ \bibnamefont
  {Greer}},\ }\href@noop {} {\ }\Eprint {http://arxiv.org/abs/arXiv:2012.11311}
  {arXiv:2012.11311} \BibitemShut {NoStop}%
\bibitem [{\citenamefont {Sau}\ and\ \citenamefont {Cohen}(2007)}]{Sau2007}%
  \BibitemOpen
  \bibfield  {author} {\bibinfo {author} {\bibfnamefont {J.~D.}\ \bibnamefont
  {Sau}}\ and\ \bibinfo {author} {\bibfnamefont {M.~L.}\ \bibnamefont
  {Cohen}},\ }\href {\doibase 10.1103/physrevb.75.045208} {\bibfield  {journal}
  {\bibinfo  {journal} {Physical Review B}\ }\textbf {\bibinfo {volume} {75}}
  (\bibinfo {year} {2007}),\ 10.1103/physrevb.75.045208}\BibitemShut {NoStop}%
\bibitem [{\citenamefont {Kao}\ \emph {et~al.}(2012)\citenamefont {Kao},
  \citenamefont {Verhulst}, \citenamefont {Vandenberghe}, \citenamefont
  {Soree}, \citenamefont {Groeseneken},\ and\ \citenamefont {Meyer}}]{Kao2012}%
  \BibitemOpen
  \bibfield  {author} {\bibinfo {author} {\bibfnamefont {K.-H.}\ \bibnamefont
  {Kao}}, \bibinfo {author} {\bibfnamefont {A.~S.}\ \bibnamefont {Verhulst}},
  \bibinfo {author} {\bibfnamefont {W.~G.}\ \bibnamefont {Vandenberghe}},
  \bibinfo {author} {\bibfnamefont {B.}~\bibnamefont {Soree}}, \bibinfo
  {author} {\bibfnamefont {G.}~\bibnamefont {Groeseneken}}, \ and\ \bibinfo
  {author} {\bibfnamefont {K.~D.}\ \bibnamefont {Meyer}},\ }\href {\doibase
  10.1109/ted.2011.2175228} {\bibfield  {journal} {\bibinfo  {journal} {{IEEE}
  Transactions on Electron Devices}\ }\textbf {\bibinfo {volume} {59}},\
  \bibinfo {pages} {292} (\bibinfo {year} {2012})}\BibitemShut {NoStop}%
\bibitem [{\citenamefont {Schulte-Braucks}\ \emph {et~al.}(2015)\citenamefont
  {Schulte-Braucks}, \citenamefont {Stange}, \citenamefont {von~den Driesch},
  \citenamefont {Blaeser}, \citenamefont {Ikonic}, \citenamefont {Hartmann},
  \citenamefont {Mantl},\ and\ \citenamefont {Buca}}]{Schulte-Braucks2015}%
  \BibitemOpen
  \bibfield  {author} {\bibinfo {author} {\bibfnamefont {C.}~\bibnamefont
  {Schulte-Braucks}}, \bibinfo {author} {\bibfnamefont {D.}~\bibnamefont
  {Stange}}, \bibinfo {author} {\bibfnamefont {N.}~\bibnamefont {von~den
  Driesch}}, \bibinfo {author} {\bibfnamefont {S.}~\bibnamefont {Blaeser}},
  \bibinfo {author} {\bibfnamefont {Z.}~\bibnamefont {Ikonic}}, \bibinfo
  {author} {\bibfnamefont {J.~M.}\ \bibnamefont {Hartmann}}, \bibinfo {author}
  {\bibfnamefont {S.}~\bibnamefont {Mantl}}, \ and\ \bibinfo {author}
  {\bibfnamefont {D.}~\bibnamefont {Buca}},\ }\href {\doibase
  10.1063/1.4927622} {\bibfield  {journal} {\bibinfo  {journal} {Applied
  Physics Letters}\ }\textbf {\bibinfo {volume} {107}},\ \bibinfo {pages}
  {042101} (\bibinfo {year} {2015})}\BibitemShut {NoStop}%
\bibitem [{\citenamefont {Ansari}\ \emph {et~al.}(2012)\citenamefont {Ansari},
  \citenamefont {Fagas}, \citenamefont {Colinge},\ and\ \citenamefont
  {Greer}}]{Ansari2012}%
  \BibitemOpen
  \bibfield  {author} {\bibinfo {author} {\bibfnamefont {L.}~\bibnamefont
  {Ansari}}, \bibinfo {author} {\bibfnamefont {G.}~\bibnamefont {Fagas}},
  \bibinfo {author} {\bibfnamefont {J.-P.}\ \bibnamefont {Colinge}}, \ and\
  \bibinfo {author} {\bibfnamefont {J.~C.}\ \bibnamefont {Greer}},\ }\href
  {\doibase 10.1021/nl2040817} {\bibfield  {journal} {\bibinfo  {journal} {Nano
  Letters}\ }\textbf {\bibinfo {volume} {12}},\ \bibinfo {pages} {2222}
  (\bibinfo {year} {2012})}\BibitemShut {NoStop}%
\bibitem [{\citenamefont {Sanchez-Soares}\ and\ \citenamefont
  {Greer}(2016)}]{Sanchez-Soares2016}%
  \BibitemOpen
  \bibfield  {author} {\bibinfo {author} {\bibfnamefont {A.}~\bibnamefont
  {Sanchez-Soares}}\ and\ \bibinfo {author} {\bibfnamefont {J.~C.}\
  \bibnamefont {Greer}},\ }\href {\doibase 10.1021/acs.nanolett.6b03612}
  {\bibfield  {journal} {\bibinfo  {journal} {Nano Letters}\ }\textbf {\bibinfo
  {volume} {16}},\ \bibinfo {pages} {7639} (\bibinfo {year}
  {2016})}\BibitemShut {NoStop}%
\bibitem [{\citenamefont {Ansari}\ \emph {et~al.}(2016)\citenamefont {Ansari},
  \citenamefont {Fagas}, \citenamefont {Gity},\ and\ \citenamefont
  {Greer}}]{Ansari2016}%
  \BibitemOpen
  \bibfield  {author} {\bibinfo {author} {\bibfnamefont {L.}~\bibnamefont
  {Ansari}}, \bibinfo {author} {\bibfnamefont {G.}~\bibnamefont {Fagas}},
  \bibinfo {author} {\bibfnamefont {F.}~\bibnamefont {Gity}}, \ and\ \bibinfo
  {author} {\bibfnamefont {J.~C.}\ \bibnamefont {Greer}},\ }\href {\doibase
  10.1063/1.4960709} {\bibfield  {journal} {\bibinfo  {journal} {Applied
  Physics Letters}\ }\textbf {\bibinfo {volume} {109}},\ \bibinfo {pages}
  {063108} (\bibinfo {year} {2016})}\BibitemShut {NoStop}%
\bibitem [{\citenamefont {Sanchez-Soares}\ \emph {et~al.}(2016)\citenamefont
  {Sanchez-Soares}, \citenamefont {O'Donnell},\ and\ \citenamefont
  {Greer}}]{Sanchez-Soares2016a}%
  \BibitemOpen
  \bibfield  {author} {\bibinfo {author} {\bibfnamefont {A.}~\bibnamefont
  {Sanchez-Soares}}, \bibinfo {author} {\bibfnamefont {C.}~\bibnamefont
  {O'Donnell}}, \ and\ \bibinfo {author} {\bibfnamefont {J.~C.}\ \bibnamefont
  {Greer}},\ }\href {\doibase 10.1103/physrevb.94.235442} {\bibfield  {journal}
  {\bibinfo  {journal} {Physical Review B}\ }\textbf {\bibinfo {volume} {94}}
  (\bibinfo {year} {2016}),\ 10.1103/physrevb.94.235442}\BibitemShut {NoStop}%
\bibitem [{\citenamefont {Gity}\ \emph {et~al.}(2017)\citenamefont {Gity},
  \citenamefont {Ansari}, \citenamefont {Lanius}, \citenamefont
  {Sch{\"u}ffelgen}, \citenamefont {Mussler}, \citenamefont {Gr{\"u}tzmacher},\
  and\ \citenamefont {Greer}}]{Gity2017}%
  \BibitemOpen
  \bibfield  {author} {\bibinfo {author} {\bibfnamefont {F.}~\bibnamefont
  {Gity}}, \bibinfo {author} {\bibfnamefont {L.}~\bibnamefont {Ansari}},
  \bibinfo {author} {\bibfnamefont {M.}~\bibnamefont {Lanius}}, \bibinfo
  {author} {\bibfnamefont {P.}~\bibnamefont {Sch{\"u}ffelgen}}, \bibinfo
  {author} {\bibfnamefont {G.}~\bibnamefont {Mussler}}, \bibinfo {author}
  {\bibfnamefont {D.}~\bibnamefont {Gr{\"u}tzmacher}}, \ and\ \bibinfo {author}
  {\bibfnamefont {J.~C.}\ \bibnamefont {Greer}},\ }\href {\doibase
  10.1063/1.4977431} {\bibfield  {journal} {\bibinfo  {journal} {Applied
  Physics Letters}\ }\textbf {\bibinfo {volume} {110}},\ \bibinfo {pages}
  {093111} (\bibinfo {year} {2017})}\BibitemShut {NoStop}%
\bibitem [{\citenamefont {Gity}\ \emph {et~al.}(2018)\citenamefont {Gity},
  \citenamefont {Ansari}, \citenamefont {K{\"o}nig}, \citenamefont {Verni},
  \citenamefont {Holmes}, \citenamefont {Long}, \citenamefont {Lanius},
  \citenamefont {Sch{\"u}ffelgen}, \citenamefont {Mussler}, \citenamefont
  {Gr{\"u}tzmacher} \emph {et~al.}}]{Gity2018}%
  \BibitemOpen
  \bibfield  {author} {\bibinfo {author} {\bibfnamefont {F.}~\bibnamefont
  {Gity}}, \bibinfo {author} {\bibfnamefont {L.}~\bibnamefont {Ansari}},
  \bibinfo {author} {\bibfnamefont {C.}~\bibnamefont {K{\"o}nig}}, \bibinfo
  {author} {\bibfnamefont {G.~A.}\ \bibnamefont {Verni}}, \bibinfo {author}
  {\bibfnamefont {J.~D.}\ \bibnamefont {Holmes}}, \bibinfo {author}
  {\bibfnamefont {B.}~\bibnamefont {Long}}, \bibinfo {author} {\bibfnamefont
  {M.}~\bibnamefont {Lanius}}, \bibinfo {author} {\bibfnamefont
  {P.}~\bibnamefont {Sch{\"u}ffelgen}}, \bibinfo {author} {\bibfnamefont
  {G.}~\bibnamefont {Mussler}}, \bibinfo {author} {\bibfnamefont
  {D.}~\bibnamefont {Gr{\"u}tzmacher}},  \emph {et~al.},\ }\href {\doibase
  10.1016/j.mee.2018.03.022} {\bibfield  {journal} {\bibinfo  {journal}
  {Microelectronic Engineering}\ }\textbf {\bibinfo {volume} {195}},\ \bibinfo
  {pages} {21} (\bibinfo {year} {2018})}\BibitemShut {NoStop}%
\bibitem [{\citenamefont {Suzuki}\ \emph {et~al.}(2016)\citenamefont {Suzuki},
  \citenamefont {Nakatsuka}, \citenamefont {Shibayama}, \citenamefont
  {Sakashita}, \citenamefont {Takeuchi}, \citenamefont {Kurosawa},\ and\
  \citenamefont {Zaima}}]{Suzuki2016}%
  \BibitemOpen
  \bibfield  {author} {\bibinfo {author} {\bibfnamefont {A.}~\bibnamefont
  {Suzuki}}, \bibinfo {author} {\bibfnamefont {O.}~\bibnamefont {Nakatsuka}},
  \bibinfo {author} {\bibfnamefont {S.}~\bibnamefont {Shibayama}}, \bibinfo
  {author} {\bibfnamefont {M.}~\bibnamefont {Sakashita}}, \bibinfo {author}
  {\bibfnamefont {W.}~\bibnamefont {Takeuchi}}, \bibinfo {author}
  {\bibfnamefont {M.}~\bibnamefont {Kurosawa}}, \ and\ \bibinfo {author}
  {\bibfnamefont {S.}~\bibnamefont {Zaima}},\ }\href {\doibase
  10.7567/jjap.55.04eb12} {\bibfield  {journal} {\bibinfo  {journal} {Japanese
  Journal of Applied Physics}\ }\textbf {\bibinfo {volume} {55}},\ \bibinfo
  {pages} {04EB12} (\bibinfo {year} {2016})}\BibitemShut {NoStop}%
\bibitem [{\citenamefont {Farrow}\ \emph {et~al.}(1981)\citenamefont {Farrow},
  \citenamefont {Robertson}, \citenamefont {Williams}, \citenamefont {Cullis},
  \citenamefont {Jones}, \citenamefont {Young},\ and\ \citenamefont
  {Dennis}}]{Farrow1981}%
  \BibitemOpen
  \bibfield  {author} {\bibinfo {author} {\bibfnamefont {R.}~\bibnamefont
  {Farrow}}, \bibinfo {author} {\bibfnamefont {D.}~\bibnamefont {Robertson}},
  \bibinfo {author} {\bibfnamefont {G.}~\bibnamefont {Williams}}, \bibinfo
  {author} {\bibfnamefont {A.}~\bibnamefont {Cullis}}, \bibinfo {author}
  {\bibfnamefont {G.}~\bibnamefont {Jones}}, \bibinfo {author} {\bibfnamefont
  {I.}~\bibnamefont {Young}}, \ and\ \bibinfo {author} {\bibfnamefont
  {P.}~\bibnamefont {Dennis}},\ }\href {\doibase 10.1016/0022-0248(81)90506-6}
  {\bibfield  {journal} {\bibinfo  {journal} {Journal of Crystal Growth}\
  }\textbf {\bibinfo {volume} {54}},\ \bibinfo {pages} {507} (\bibinfo {year}
  {1981})}\BibitemShut {NoStop}%
\bibitem [{\citenamefont {Hochst}\ and\ \citenamefont
  {Hernandez-Calderon}(1983)}]{Hoechst1983}%
  \BibitemOpen
  \bibfield  {author} {\bibinfo {author} {\bibfnamefont {H.}~\bibnamefont
  {Hochst}}\ and\ \bibinfo {author} {\bibfnamefont {I.}~\bibnamefont
  {Hernandez-Calderon}},\ }\href {\doibase 10.1016/0039-6028(83)90691-X}
  {\bibfield  {journal} {\bibinfo  {journal} {Surface Science}\ }\textbf
  {\bibinfo {volume} {126}},\ \bibinfo {pages} {25} (\bibinfo {year}
  {1983})}\BibitemShut {NoStop}%
\bibitem [{\citenamefont {Asom}\ \emph {et~al.}(1989)\citenamefont {Asom},
  \citenamefont {Fitzgerald}, \citenamefont {Kortan}, \citenamefont {Spear},\
  and\ \citenamefont {Kimerling}}]{Asom1989}%
  \BibitemOpen
  \bibfield  {author} {\bibinfo {author} {\bibfnamefont {M.~T.}\ \bibnamefont
  {Asom}}, \bibinfo {author} {\bibfnamefont {E.~A.}\ \bibnamefont
  {Fitzgerald}}, \bibinfo {author} {\bibfnamefont {A.~R.}\ \bibnamefont
  {Kortan}}, \bibinfo {author} {\bibfnamefont {B.}~\bibnamefont {Spear}}, \
  and\ \bibinfo {author} {\bibfnamefont {L.~C.}\ \bibnamefont {Kimerling}},\
  }\href {\doibase 10.1063/1.101838} {\bibfield  {journal} {\bibinfo  {journal}
  {Applied Physics Letters}\ }\textbf {\bibinfo {volume} {55}},\ \bibinfo
  {pages} {578} (\bibinfo {year} {1989})}\BibitemShut {NoStop}%
\bibitem [{\citenamefont {John}\ \emph {et~al.}(1989)\citenamefont {John},
  \citenamefont {Miller},\ and\ \citenamefont {Chiang}}]{John1989}%
  \BibitemOpen
  \bibfield  {author} {\bibinfo {author} {\bibfnamefont {P.}~\bibnamefont
  {John}}, \bibinfo {author} {\bibfnamefont {T.}~\bibnamefont {Miller}}, \ and\
  \bibinfo {author} {\bibfnamefont {T.-C.}\ \bibnamefont {Chiang}},\ }\href
  {\doibase 10.1103/PhysRevB.39.3223} {\bibfield  {journal} {\bibinfo
  {journal} {Physical Review B}\ }\textbf {\bibinfo {volume} {39}},\ \bibinfo
  {pages} {3223} (\bibinfo {year} {1989})}\BibitemShut {NoStop}%
\bibitem [{\citenamefont {Bowman}\ \emph {et~al.}(1990)\citenamefont {Bowman},
  \citenamefont {Adams}, \citenamefont {Engelhart},\ and\ \citenamefont
  {Höchst}}]{Bowman1990}%
  \BibitemOpen
  \bibfield  {author} {\bibinfo {author} {\bibfnamefont {R.~C.}\ \bibnamefont
  {Bowman}}, \bibinfo {author} {\bibfnamefont {P.~M.}\ \bibnamefont {Adams}},
  \bibinfo {author} {\bibfnamefont {M.~A.}\ \bibnamefont {Engelhart}}, \ and\
  \bibinfo {author} {\bibfnamefont {H.}~\bibnamefont {Höchst}},\ }\href
  {\doibase 10.1116/1.576768} {\bibfield  {journal} {\bibinfo  {journal}
  {Journal of Vacuum Science {\&} Technology A: Vacuum, Surfaces, and Films}\
  }\textbf {\bibinfo {volume} {8}},\ \bibinfo {pages} {1577} (\bibinfo {year}
  {1990})}\BibitemShut {NoStop}%
\bibitem [{\citenamefont {Bragg}\ and\ \citenamefont
  {Williams}(1934)}]{Bragg1934}%
  \BibitemOpen
  \bibfield  {author} {\bibinfo {author} {\bibfnamefont {W.~L.}\ \bibnamefont
  {Bragg}}\ and\ \bibinfo {author} {\bibfnamefont {E.~J.}\ \bibnamefont
  {Williams}},\ }\href {\doibase 10.1098/rspa.1934.0132} {\bibfield  {journal}
  {\bibinfo  {journal} {Proceedings of the Royal Society A: Mathematical,
  Physical and Engineering Sciences}\ }\textbf {\bibinfo {volume} {145}},\
  \bibinfo {pages} {699} (\bibinfo {year} {1934})}\BibitemShut {NoStop}%
\bibitem [{\citenamefont {Williams}(1935{\natexlab{a}})}]{WilliamsII1935}%
  \BibitemOpen
  \bibfield  {author} {\bibinfo {author} {\bibfnamefont {E.}~\bibnamefont
  {Williams}},\ }\href {\doibase 10.1098/rspa.1935.0165} {\bibfield  {journal}
  {\bibinfo  {journal} {Proceedings of the Royal Society of London. Series A,
  Mathematical and Physical Sciences}\ }\textbf {\bibinfo {volume} {151}},\
  \bibinfo {pages} {231} (\bibinfo {year} {1935}{\natexlab{a}})}\BibitemShut
  {NoStop}%
\bibitem [{\citenamefont {Williams}(1935{\natexlab{b}})}]{Williams1935}%
  \BibitemOpen
  \bibfield  {author} {\bibinfo {author} {\bibfnamefont {E.}~\bibnamefont
  {Williams}},\ }\href {\doibase 10.1098/rspa.1935.0188} {\bibfield  {journal}
  {\bibinfo  {journal} {Proceedings of the Royal Society of London. Series A,
  Mathematical and Physical Sciences}\ }\textbf {\bibinfo {volume} {152}},\
  \bibinfo {pages} {231} (\bibinfo {year} {1935}{\natexlab{b}})}\BibitemShut
  {NoStop}%
\bibitem [{\citenamefont {Sanchez}\ \emph {et~al.}(1984)\citenamefont
  {Sanchez}, \citenamefont {Ducastelle},\ and\ \citenamefont
  {Gratias}}]{Sanchez1984}%
  \BibitemOpen
  \bibfield  {author} {\bibinfo {author} {\bibfnamefont {J.~M.}\ \bibnamefont
  {Sanchez}}, \bibinfo {author} {\bibfnamefont {F.}~\bibnamefont {Ducastelle}},
  \ and\ \bibinfo {author} {\bibfnamefont {D.}~\bibnamefont {Gratias}},\ }\href
  {\doibase 10.1016/0378-4371(84)90096-7} {\bibfield  {journal} {\bibinfo
  {journal} {Physica A. Statistical and Theoretical Physics}\ }\textbf
  {\bibinfo {volume} {128}},\ \bibinfo {pages} {334} (\bibinfo {year}
  {1984})}\BibitemShut {NoStop}%
\bibitem [{\citenamefont {Soler}\ \emph {et~al.}(2002)\citenamefont {Soler},
  \citenamefont {Artacho}, \citenamefont {Gale}, \citenamefont {Garc{\'{\i}}a},
  \citenamefont {Junquera}, \citenamefont {Ordej{\'{o}}n},\ and\ \citenamefont
  {S{\'{a}}nchez-Portal}}]{Soler2002}%
  \BibitemOpen
  \bibfield  {author} {\bibinfo {author} {\bibfnamefont {J.~M.}\ \bibnamefont
  {Soler}}, \bibinfo {author} {\bibfnamefont {E.}~\bibnamefont {Artacho}},
  \bibinfo {author} {\bibfnamefont {J.~D.}\ \bibnamefont {Gale}}, \bibinfo
  {author} {\bibfnamefont {A.}~\bibnamefont {Garc{\'{\i}}a}}, \bibinfo {author}
  {\bibfnamefont {J.}~\bibnamefont {Junquera}}, \bibinfo {author}
  {\bibfnamefont {P.}~\bibnamefont {Ordej{\'{o}}n}}, \ and\ \bibinfo {author}
  {\bibfnamefont {D.}~\bibnamefont {S{\'{a}}nchez-Portal}},\ }\href {\doibase
  10.1088/0953-8984/14/11/302} {\bibfield  {journal} {\bibinfo  {journal}
  {Journal of Physics: Condensed Matter}\ }\textbf {\bibinfo {volume} {14}},\
  \bibinfo {pages} {2745} (\bibinfo {year} {2002})}\BibitemShut {NoStop}%
\bibitem [{\citenamefont {Ozaki}(2003)}]{Ozaki03}%
  \BibitemOpen
  \bibfield  {author} {\bibinfo {author} {\bibfnamefont {T.}~\bibnamefont
  {Ozaki}},\ }\href {\doibase 10.1103/PhysRevB.67.155108} {\bibfield  {journal}
  {\bibinfo  {journal} {Phys. Rev. B}\ }\textbf {\bibinfo {volume} {67}},\
  \bibinfo {pages} {155108} (\bibinfo {year} {2003})}\BibitemShut {NoStop}%
\bibitem [{\citenamefont {Ozaki}\ and\ \citenamefont {Kino}(2004)}]{Ozaki04}%
  \BibitemOpen
  \bibfield  {author} {\bibinfo {author} {\bibfnamefont {T.}~\bibnamefont
  {Ozaki}}\ and\ \bibinfo {author} {\bibfnamefont {H.}~\bibnamefont {Kino}},\
  }\href {\doibase 10.1103/PhysRevB.69.195113} {\bibfield  {journal} {\bibinfo
  {journal} {Phys. Rev. B}\ }\textbf {\bibinfo {volume} {69}},\ \bibinfo
  {pages} {195113} (\bibinfo {year} {2004})}\BibitemShut {NoStop}%
\bibitem [{\citenamefont {{Atomistix Toolkit version 2016.4}}()}]{QW}%
  \BibitemOpen
  \bibfield  {author} {\bibinfo {author} {\bibnamefont {{Atomistix Toolkit
  version 2016.4}}},\ }\href@noop {} {\enquote {\bibinfo {title} {{QuantumWise
  A/S (www.quantumwise.com).}}}\ }\BibitemShut {NoStop}%
\bibitem [{\citenamefont {Monkhorst}\ and\ \citenamefont
  {Pack}(1976)}]{Monkhorst1976}%
  \BibitemOpen
  \bibfield  {author} {\bibinfo {author} {\bibfnamefont {H.~J.}\ \bibnamefont
  {Monkhorst}}\ and\ \bibinfo {author} {\bibfnamefont {J.~D.}\ \bibnamefont
  {Pack}},\ }\href {\doibase 10.1103/PhysRevB.13.5188} {\bibfield  {journal}
  {\bibinfo  {journal} {Phys. Rev. B}\ }\textbf {\bibinfo {volume} {13}},\
  \bibinfo {pages} {5188} (\bibinfo {year} {1976})}\BibitemShut {NoStop}%
\bibitem [{\citenamefont {Perdew}\ and\ \citenamefont
  {Zunger}(1981)}]{Perdew1981}%
  \BibitemOpen
  \bibfield  {author} {\bibinfo {author} {\bibfnamefont {J.~P.}\ \bibnamefont
  {Perdew}}\ and\ \bibinfo {author} {\bibfnamefont {A.}~\bibnamefont
  {Zunger}},\ }\href {\doibase 10.1103/physrevb.23.5048} {\bibfield  {journal}
  {\bibinfo  {journal} {Physical Review B}\ }\textbf {\bibinfo {volume} {23}},\
  \bibinfo {pages} {5048} (\bibinfo {year} {1981})}\BibitemShut {NoStop}%
\bibitem [{\citenamefont {Baker}\ and\ \citenamefont {Hart}(1975)}]{Baker1975}%
  \BibitemOpen
  \bibfield  {author} {\bibinfo {author} {\bibfnamefont {J.~F.~C.}\
  \bibnamefont {Baker}}\ and\ \bibinfo {author} {\bibfnamefont
  {M.}~\bibnamefont {Hart}},\ }\href {\doibase 10.1107/s0567739475000769}
  {\bibfield  {journal} {\bibinfo  {journal} {Acta Crystallographica Section
  A}\ }\textbf {\bibinfo {volume} {31}},\ \bibinfo {pages} {364} (\bibinfo
  {year} {1975})}\BibitemShut {NoStop}%
\bibitem [{\citenamefont {Thewlis}\ and\ \citenamefont
  {Davey}(1954)}]{Thewlis1954}%
  \BibitemOpen
  \bibfield  {author} {\bibinfo {author} {\bibfnamefont {J.}~\bibnamefont
  {Thewlis}}\ and\ \bibinfo {author} {\bibfnamefont {A.}~\bibnamefont
  {Davey}},\ }\href {\doibase 10.1038/1741011a0} {\bibfield  {journal}
  {\bibinfo  {journal} {Nature}\ }\textbf {\bibinfo {volume} {174}},\ \bibinfo
  {pages} {1011} (\bibinfo {year} {1954})}\BibitemShut {NoStop}%
\bibitem [{\citenamefont {Bruner}\ and\ \citenamefont
  {Keyes}(1961)}]{Bruner1961}%
  \BibitemOpen
  \bibfield  {author} {\bibinfo {author} {\bibfnamefont {L.~J.}\ \bibnamefont
  {Bruner}}\ and\ \bibinfo {author} {\bibfnamefont {R.~W.}\ \bibnamefont
  {Keyes}},\ }\href {\doibase 10.1103/physrevlett.7.55} {\bibfield  {journal}
  {\bibinfo  {journal} {Physical Review Letters}\ }\textbf {\bibinfo {volume}
  {7}},\ \bibinfo {pages} {55} (\bibinfo {year} {1961})}\BibitemShut {NoStop}%
\bibitem [{\citenamefont {Price}\ \emph {et~al.}(1971)\citenamefont {Price},
  \citenamefont {Rowe},\ and\ \citenamefont {Nicklow}}]{Price1971}%
  \BibitemOpen
  \bibfield  {author} {\bibinfo {author} {\bibfnamefont {D.~L.}\ \bibnamefont
  {Price}}, \bibinfo {author} {\bibfnamefont {J.~M.}\ \bibnamefont {Rowe}}, \
  and\ \bibinfo {author} {\bibfnamefont {R.~M.}\ \bibnamefont {Nicklow}},\
  }\href {\doibase 10.1103/physrevb.3.1268} {\bibfield  {journal} {\bibinfo
  {journal} {Physical Review B}\ }\textbf {\bibinfo {volume} {3}},\ \bibinfo
  {pages} {1268} (\bibinfo {year} {1971})}\BibitemShut {NoStop}%
\bibitem [{\citenamefont {Buchenauer}\ \emph {et~al.}(1971)\citenamefont
  {Buchenauer}, \citenamefont {Cardona},\ and\ \citenamefont
  {Pollak}}]{Buchenauer1971}%
  \BibitemOpen
  \bibfield  {author} {\bibinfo {author} {\bibfnamefont {C.}~\bibnamefont
  {Buchenauer}}, \bibinfo {author} {\bibfnamefont {M.}~\bibnamefont {Cardona}},
  \ and\ \bibinfo {author} {\bibfnamefont {F.}~\bibnamefont {Pollak}},\ }\href
  {\doibase 10.1103/PhysRevB.3.1243} {\bibfield  {journal} {\bibinfo  {journal}
  {Physical Review B}\ }\textbf {\bibinfo {volume} {3}},\ \bibinfo {pages}
  {1243} (\bibinfo {year} {1971})}\BibitemShut {NoStop}%
\bibitem [{\citenamefont {Biswas}\ \emph {et~al.}(2016)\citenamefont {Biswas},
  \citenamefont {Doherty}, \citenamefont {Saladukha}, \citenamefont {Ramasse},
  \citenamefont {Majumdar}, \citenamefont {Upmanyu}, \citenamefont {Singha},
  \citenamefont {Ochalski}, \citenamefont {Morris},\ and\ \citenamefont
  {Holmes}}]{Biswas2016}%
  \BibitemOpen
  \bibfield  {author} {\bibinfo {author} {\bibfnamefont {S.}~\bibnamefont
  {Biswas}}, \bibinfo {author} {\bibfnamefont {J.}~\bibnamefont {Doherty}},
  \bibinfo {author} {\bibfnamefont {D.}~\bibnamefont {Saladukha}}, \bibinfo
  {author} {\bibfnamefont {Q.}~\bibnamefont {Ramasse}}, \bibinfo {author}
  {\bibfnamefont {D.}~\bibnamefont {Majumdar}}, \bibinfo {author}
  {\bibfnamefont {M.}~\bibnamefont {Upmanyu}}, \bibinfo {author} {\bibfnamefont
  {A.}~\bibnamefont {Singha}}, \bibinfo {author} {\bibfnamefont
  {T.}~\bibnamefont {Ochalski}}, \bibinfo {author} {\bibfnamefont {M.~A.}\
  \bibnamefont {Morris}}, \ and\ \bibinfo {author} {\bibfnamefont {J.~D.}\
  \bibnamefont {Holmes}},\ }\href {\doibase 10.1038/ncomms11405} {\bibfield
  {journal} {\bibinfo  {journal} {Nature Communications}\ }\textbf {\bibinfo
  {volume} {7}} (\bibinfo {year} {2016}),\ 10.1038/ncomms11405}\BibitemShut
  {NoStop}%
\bibitem [{\citenamefont {Mukherjee}\ \emph {et~al.}(2017)\citenamefont
  {Mukherjee}, \citenamefont {Kodali}, \citenamefont {Isheim}, \citenamefont
  {Wirths}, \citenamefont {Hartmann}, \citenamefont {Buca}, \citenamefont
  {Seidman},\ and\ \citenamefont {Moutanabbir}}]{Mukherjee2017}%
  \BibitemOpen
  \bibfield  {author} {\bibinfo {author} {\bibfnamefont {S.}~\bibnamefont
  {Mukherjee}}, \bibinfo {author} {\bibfnamefont {N.}~\bibnamefont {Kodali}},
  \bibinfo {author} {\bibfnamefont {D.}~\bibnamefont {Isheim}}, \bibinfo
  {author} {\bibfnamefont {S.}~\bibnamefont {Wirths}}, \bibinfo {author}
  {\bibfnamefont {J.~M.}\ \bibnamefont {Hartmann}}, \bibinfo {author}
  {\bibfnamefont {D.}~\bibnamefont {Buca}}, \bibinfo {author} {\bibfnamefont
  {D.~N.}\ \bibnamefont {Seidman}}, \ and\ \bibinfo {author} {\bibfnamefont
  {O.}~\bibnamefont {Moutanabbir}},\ }\href {\doibase
  10.1103/physrevb.95.161402} {\bibfield  {journal} {\bibinfo  {journal}
  {Physical Review B}\ }\textbf {\bibinfo {volume} {95}} (\bibinfo {year}
  {2017}),\ 10.1103/physrevb.95.161402}\BibitemShut {NoStop}%
\bibitem [{\citenamefont {Connolly}\ and\ \citenamefont
  {Williams}(1983)}]{Connolly1983}%
  \BibitemOpen
  \bibfield  {author} {\bibinfo {author} {\bibfnamefont {J.~W.~D.}\
  \bibnamefont {Connolly}}\ and\ \bibinfo {author} {\bibfnamefont {A.~R.}\
  \bibnamefont {Williams}},\ }\href {\doibase 10.1103/physrevb.27.5169}
  {\bibfield  {journal} {\bibinfo  {journal} {Physical Review B}\ }\textbf
  {\bibinfo {volume} {27}},\ \bibinfo {pages} {5169} (\bibinfo {year}
  {1983})}\BibitemShut {NoStop}%
\bibitem [{\citenamefont {van~de Walle}\ and\ \citenamefont
  {Ceder}(2002{\natexlab{a}})}]{Walle2002}%
  \BibitemOpen
  \bibfield  {author} {\bibinfo {author} {\bibfnamefont {A.}~\bibnamefont
  {van~de Walle}}\ and\ \bibinfo {author} {\bibfnamefont {G.}~\bibnamefont
  {Ceder}},\ }\href {\doibase 10.1361/105497102770331596} {\bibfield  {journal}
  {\bibinfo  {journal} {Journal of Phase Equilibria}\ }\textbf {\bibinfo
  {volume} {23}},\ \bibinfo {pages} {348} (\bibinfo {year}
  {2002}{\natexlab{a}})}\BibitemShut {NoStop}%
\bibitem [{\citenamefont {Stone}(1974)}]{Stone1974}%
  \BibitemOpen
  \bibfield  {author} {\bibinfo {author} {\bibfnamefont {M.}~\bibnamefont
  {Stone}},\ }\href {\doibase 10.1111/j.2517-6161.1974.tb00994.x} {\bibfield
  {journal} {\bibinfo  {journal} {Journal of the Royal Statistical Society:
  Series B (Methodological)}\ }\textbf {\bibinfo {volume} {36}},\ \bibinfo
  {pages} {111} (\bibinfo {year} {1974})}\BibitemShut {NoStop}%
\bibitem [{\citenamefont {Wolverton}\ and\ \citenamefont
  {Zunger}(1995)}]{Wolverton1995}%
  \BibitemOpen
  \bibfield  {author} {\bibinfo {author} {\bibfnamefont {C.}~\bibnamefont
  {Wolverton}}\ and\ \bibinfo {author} {\bibfnamefont {A.}~\bibnamefont
  {Zunger}},\ }\href {\doibase 10.1103/physrevb.52.8813} {\bibfield  {journal}
  {\bibinfo  {journal} {Physical Review B}\ }\textbf {\bibinfo {volume} {52}},\
  \bibinfo {pages} {8813} (\bibinfo {year} {1995})}\BibitemShut {NoStop}%
\bibitem [{\citenamefont {van~de Walle}(2009)}]{Walle2009}%
  \BibitemOpen
  \bibfield  {author} {\bibinfo {author} {\bibfnamefont {A.}~\bibnamefont
  {van~de Walle}},\ }\href {\doibase 10.1016/j.calphad.2008.12.005} {\bibfield
  {journal} {\bibinfo  {journal} {Calphad}\ }\textbf {\bibinfo {volume} {33}},\
  \bibinfo {pages} {266} (\bibinfo {year} {2009})}\BibitemShut {NoStop}%
\bibitem [{\citenamefont {Zunger}\ \emph {et~al.}(1990)\citenamefont {Zunger},
  \citenamefont {Wei}, \citenamefont {Ferreira},\ and\ \citenamefont
  {Bernard}}]{Zunger1990}%
  \BibitemOpen
  \bibfield  {author} {\bibinfo {author} {\bibfnamefont {A.}~\bibnamefont
  {Zunger}}, \bibinfo {author} {\bibfnamefont {S.-H.}\ \bibnamefont {Wei}},
  \bibinfo {author} {\bibfnamefont {L.}~\bibnamefont {Ferreira}}, \ and\
  \bibinfo {author} {\bibfnamefont {J.~E.}\ \bibnamefont {Bernard}},\ }\href
  {\doibase 10.1103/PhysRevLett.65.353} {\bibfield  {journal} {\bibinfo
  {journal} {Physical Review Letters}\ }\textbf {\bibinfo {volume} {65}},\
  \bibinfo {pages} {353} (\bibinfo {year} {1990})}\BibitemShut {NoStop}%
\bibitem [{\citenamefont {Hass}\ \emph {et~al.}(1990)\citenamefont {Hass},
  \citenamefont {Davis},\ and\ \citenamefont {Zunger}}]{Hass1990}%
  \BibitemOpen
  \bibfield  {author} {\bibinfo {author} {\bibfnamefont {K.}~\bibnamefont
  {Hass}}, \bibinfo {author} {\bibfnamefont {L.}~\bibnamefont {Davis}}, \ and\
  \bibinfo {author} {\bibfnamefont {A.}~\bibnamefont {Zunger}},\ }\href
  {\doibase 10.1103/PhysRevB.42.3757} {\bibfield  {journal} {\bibinfo
  {journal} {Physical Review B}\ }\textbf {\bibinfo {volume} {42}},\ \bibinfo
  {pages} {3757} (\bibinfo {year} {1990})}\BibitemShut {NoStop}%
\bibitem [{\citenamefont {Van~de Walle}\ \emph {et~al.}(2013)\citenamefont
  {Van~de Walle}, \citenamefont {Tiwary}, \citenamefont {De~Jong},
  \citenamefont {Olmsted}, \citenamefont {Asta}, \citenamefont {Dick},
  \citenamefont {Shin}, \citenamefont {Wang}, \citenamefont {Chen},\ and\
  \citenamefont {Liu}}]{VandeWalle2013}%
  \BibitemOpen
  \bibfield  {author} {\bibinfo {author} {\bibfnamefont {A.}~\bibnamefont
  {Van~de Walle}}, \bibinfo {author} {\bibfnamefont {P.}~\bibnamefont
  {Tiwary}}, \bibinfo {author} {\bibfnamefont {M.}~\bibnamefont {De~Jong}},
  \bibinfo {author} {\bibfnamefont {D.}~\bibnamefont {Olmsted}}, \bibinfo
  {author} {\bibfnamefont {M.}~\bibnamefont {Asta}}, \bibinfo {author}
  {\bibfnamefont {A.}~\bibnamefont {Dick}}, \bibinfo {author} {\bibfnamefont
  {D.}~\bibnamefont {Shin}}, \bibinfo {author} {\bibfnamefont {Y.}~\bibnamefont
  {Wang}}, \bibinfo {author} {\bibfnamefont {L.-Q.}\ \bibnamefont {Chen}}, \
  and\ \bibinfo {author} {\bibfnamefont {Z.-K.}\ \bibnamefont {Liu}},\ }\href
  {\doibase 10.1016/j.calphad.2013.06.006} {\bibfield  {journal} {\bibinfo
  {journal} {Calphad}\ }\textbf {\bibinfo {volume} {42}},\ \bibinfo {pages}
  {13} (\bibinfo {year} {2013})}\BibitemShut {NoStop}%
\bibitem [{\citenamefont {van~de Walle}\ and\ \citenamefont
  {Ceder}(2002{\natexlab{b}})}]{Walle2002a}%
  \BibitemOpen
  \bibfield  {author} {\bibinfo {author} {\bibfnamefont {A.}~\bibnamefont
  {van~de Walle}}\ and\ \bibinfo {author} {\bibfnamefont {G.}~\bibnamefont
  {Ceder}},\ }\href {\doibase 10.1103/revmodphys.74.11} {\bibfield  {journal}
  {\bibinfo  {journal} {Reviews of Modern Physics}\ }\textbf {\bibinfo {volume}
  {74}},\ \bibinfo {pages} {11} (\bibinfo {year}
  {2002}{\natexlab{b}})}\BibitemShut {NoStop}%
\bibitem [{\citenamefont {Fultz}(2010)}]{Fultz2010}%
  \BibitemOpen
  \bibfield  {author} {\bibinfo {author} {\bibfnamefont {B.}~\bibnamefont
  {Fultz}},\ }\href {\doibase 10.1016/j.pmatsci.2009.05.002} {\bibfield
  {journal} {\bibinfo  {journal} {Progress in Materials Science}\ }\textbf
  {\bibinfo {volume} {55}},\ \bibinfo {pages} {247} (\bibinfo {year}
  {2010})}\BibitemShut {NoStop}%
\bibitem [{\citenamefont {Bozzolo}\ \emph {et~al.}(2007)\citenamefont
  {Bozzolo}, \citenamefont {Noebe},\ and\ \citenamefont {Abel}}]{Bozzolo2007}%
  \BibitemOpen
  \bibfield  {author} {\bibinfo {author} {\bibfnamefont {G.}~\bibnamefont
  {Bozzolo}}, \bibinfo {author} {\bibfnamefont {R.~D.}\ \bibnamefont {Noebe}},
  \ and\ \bibinfo {author} {\bibfnamefont {P.~B.}\ \bibnamefont {Abel}},\
  }\href@noop {} {\emph {\bibinfo {title} {Applied computational materials
  modeling: theory, simulation and experiment}}}\ (\bibinfo  {publisher}
  {Springer Science \& Business Media},\ \bibinfo {address} {New York},\
  \bibinfo {year} {2007})\BibitemShut {NoStop}%
\bibitem [{\citenamefont {van~de Walle}\ and\ \citenamefont
  {Asta}(2002)}]{Walle2002b}%
  \BibitemOpen
  \bibfield  {author} {\bibinfo {author} {\bibfnamefont {A.}~\bibnamefont
  {van~de Walle}}\ and\ \bibinfo {author} {\bibfnamefont {M.}~\bibnamefont
  {Asta}},\ }\href {\doibase 10.1088/0965-0393/10/5/304} {\bibfield  {journal}
  {\bibinfo  {journal} {Modelling and Simulation in Materials Science and
  Engineering}\ }\textbf {\bibinfo {volume} {10}},\ \bibinfo {pages} {521}
  (\bibinfo {year} {2002})}\BibitemShut {NoStop}%
\bibitem [{\citenamefont {Ghosh}\ \emph {et~al.}(2008)\citenamefont {Ghosh},
  \citenamefont {van~de Walle},\ and\ \citenamefont {Asta}}]{Ghosh2008}%
  \BibitemOpen
  \bibfield  {author} {\bibinfo {author} {\bibfnamefont {G.}~\bibnamefont
  {Ghosh}}, \bibinfo {author} {\bibfnamefont {A.}~\bibnamefont {van~de Walle}},
  \ and\ \bibinfo {author} {\bibfnamefont {M.}~\bibnamefont {Asta}},\ }\href
  {\doibase 10.1016/j.actamat.2008.03.006} {\bibfield  {journal} {\bibinfo
  {journal} {Acta Materialia}\ }\textbf {\bibinfo {volume} {56}},\ \bibinfo
  {pages} {3202} (\bibinfo {year} {2008})}\BibitemShut {NoStop}%
\bibitem [{\citenamefont {O'Halloran}\ \emph {et~al.}(2019)\citenamefont
  {O'Halloran}, \citenamefont {Broderick}, \citenamefont {Tanner},
  \citenamefont {Schulz},\ and\ \citenamefont {O'Reilly}}]{OHalloran2019}%
  \BibitemOpen
  \bibfield  {author} {\bibinfo {author} {\bibfnamefont {E.~J.}\ \bibnamefont
  {O'Halloran}}, \bibinfo {author} {\bibfnamefont {C.~A.}\ \bibnamefont
  {Broderick}}, \bibinfo {author} {\bibfnamefont {D.~S.~P.}\ \bibnamefont
  {Tanner}}, \bibinfo {author} {\bibfnamefont {S.}~\bibnamefont {Schulz}}, \
  and\ \bibinfo {author} {\bibfnamefont {E.~P.}\ \bibnamefont {O'Reilly}},\
  }\href {\doibase 10.1007/s11082-019-1992-8} {\bibfield  {journal} {\bibinfo
  {journal} {Optical and Quantum Electronics}\ }\textbf {\bibinfo {volume}
  {51}},\ \bibinfo {pages} {314} (\bibinfo {year} {2019})}\BibitemShut
  {NoStop}%
\bibitem [{\citenamefont {Gencarelli}\ \emph {et~al.}(2015)\citenamefont
  {Gencarelli}, \citenamefont {Grandjean}, \citenamefont {Shimura},
  \citenamefont {Vincent}, \citenamefont {Banerjee}, \citenamefont {Vantomme},
  \citenamefont {Vandervorst}, \citenamefont {Loo}, \citenamefont {Heyns},\
  and\ \citenamefont {Temst}}]{Gencarelli2015}%
  \BibitemOpen
  \bibfield  {author} {\bibinfo {author} {\bibfnamefont {F.}~\bibnamefont
  {Gencarelli}}, \bibinfo {author} {\bibfnamefont {D.}~\bibnamefont
  {Grandjean}}, \bibinfo {author} {\bibfnamefont {Y.}~\bibnamefont {Shimura}},
  \bibinfo {author} {\bibfnamefont {B.}~\bibnamefont {Vincent}}, \bibinfo
  {author} {\bibfnamefont {D.}~\bibnamefont {Banerjee}}, \bibinfo {author}
  {\bibfnamefont {A.}~\bibnamefont {Vantomme}}, \bibinfo {author}
  {\bibfnamefont {W.}~\bibnamefont {Vandervorst}}, \bibinfo {author}
  {\bibfnamefont {R.}~\bibnamefont {Loo}}, \bibinfo {author} {\bibfnamefont
  {M.}~\bibnamefont {Heyns}}, \ and\ \bibinfo {author} {\bibfnamefont
  {K.}~\bibnamefont {Temst}},\ }\href {\doibase 10.1063/1.4913856} {\bibfield
  {journal} {\bibinfo  {journal} {Journal of Applied Physics}\ }\textbf
  {\bibinfo {volume} {117}},\ \bibinfo {pages} {095702} (\bibinfo {year}
  {2015})}\BibitemShut {NoStop}%
\bibitem [{\citenamefont {Xu}\ \emph {et~al.}(2017)\citenamefont {Xu},
  \citenamefont {Senaratne}, \citenamefont {Culbertson}, \citenamefont
  {Kouvetakis},\ and\ \citenamefont {Men{\'{e}}ndez}}]{Xu2017}%
  \BibitemOpen
  \bibfield  {author} {\bibinfo {author} {\bibfnamefont {C.}~\bibnamefont
  {Xu}}, \bibinfo {author} {\bibfnamefont {C.~L.}\ \bibnamefont {Senaratne}},
  \bibinfo {author} {\bibfnamefont {R.~J.}\ \bibnamefont {Culbertson}},
  \bibinfo {author} {\bibfnamefont {J.}~\bibnamefont {Kouvetakis}}, \ and\
  \bibinfo {author} {\bibfnamefont {J.}~\bibnamefont {Men{\'{e}}ndez}},\ }\href
  {\doibase 10.1063/1.4996306} {\bibfield  {journal} {\bibinfo  {journal}
  {Journal of Applied Physics}\ }\textbf {\bibinfo {volume} {122}},\ \bibinfo
  {pages} {125702} (\bibinfo {year} {2017})}\BibitemShut {NoStop}%
\bibitem [{\citenamefont {Gassenq}\ \emph {et~al.}(2017)\citenamefont
  {Gassenq}, \citenamefont {Milord}, \citenamefont {Aubin}, \citenamefont
  {Pauc}, \citenamefont {Guilloy}, \citenamefont {Rothman}, \citenamefont
  {Rouchon}, \citenamefont {Chelnokov}, \citenamefont {Hartmann}, \citenamefont
  {Reboud},\ and\ \citenamefont {Calvo}}]{Gassenq2017}%
  \BibitemOpen
  \bibfield  {author} {\bibinfo {author} {\bibfnamefont {A.}~\bibnamefont
  {Gassenq}}, \bibinfo {author} {\bibfnamefont {L.}~\bibnamefont {Milord}},
  \bibinfo {author} {\bibfnamefont {J.}~\bibnamefont {Aubin}}, \bibinfo
  {author} {\bibfnamefont {N.}~\bibnamefont {Pauc}}, \bibinfo {author}
  {\bibfnamefont {K.}~\bibnamefont {Guilloy}}, \bibinfo {author} {\bibfnamefont
  {J.}~\bibnamefont {Rothman}}, \bibinfo {author} {\bibfnamefont
  {D.}~\bibnamefont {Rouchon}}, \bibinfo {author} {\bibfnamefont
  {A.}~\bibnamefont {Chelnokov}}, \bibinfo {author} {\bibfnamefont {J.~M.}\
  \bibnamefont {Hartmann}}, \bibinfo {author} {\bibfnamefont {V.}~\bibnamefont
  {Reboud}}, \ and\ \bibinfo {author} {\bibfnamefont {V.}~\bibnamefont
  {Calvo}},\ }\href {\doibase 10.1063/1.4978512} {\bibfield  {journal}
  {\bibinfo  {journal} {Applied Physics Letters}\ }\textbf {\bibinfo {volume}
  {110}},\ \bibinfo {pages} {112101} (\bibinfo {year} {2017})}\BibitemShut
  {NoStop}%
\bibitem [{\citenamefont {Xu}\ \emph {et~al.}(2019)\citenamefont {Xu},
  \citenamefont {Wallace}, \citenamefont {Ringwala}, \citenamefont {Chang},
  \citenamefont {Poweleit}, \citenamefont {Kouvetakis},\ and\ \citenamefont
  {Men{\'{e}}ndez}}]{Xu2019}%
  \BibitemOpen
  \bibfield  {author} {\bibinfo {author} {\bibfnamefont {C.}~\bibnamefont
  {Xu}}, \bibinfo {author} {\bibfnamefont {P.~M.}\ \bibnamefont {Wallace}},
  \bibinfo {author} {\bibfnamefont {D.~A.}\ \bibnamefont {Ringwala}}, \bibinfo
  {author} {\bibfnamefont {S.~L.~Y.}\ \bibnamefont {Chang}}, \bibinfo {author}
  {\bibfnamefont {C.~D.}\ \bibnamefont {Poweleit}}, \bibinfo {author}
  {\bibfnamefont {J.}~\bibnamefont {Kouvetakis}}, \ and\ \bibinfo {author}
  {\bibfnamefont {J.}~\bibnamefont {Men{\'{e}}ndez}},\ }\href {\doibase
  10.1063/1.5100275} {\bibfield  {journal} {\bibinfo  {journal} {Applied
  Physics Letters}\ }\textbf {\bibinfo {volume} {114}},\ \bibinfo {pages}
  {212104} (\bibinfo {year} {2019})}\BibitemShut {NoStop}%
\bibitem [{\citenamefont {Bouthillier}\ \emph {et~al.}(2020)\citenamefont
  {Bouthillier}, \citenamefont {Assali}, \citenamefont {Nicolas},\ and\
  \citenamefont {Moutanabbir}}]{Bouthillier2020}%
  \BibitemOpen
  \bibfield  {author} {\bibinfo {author} {\bibfnamefont {{\'{E}}.}~\bibnamefont
  {Bouthillier}}, \bibinfo {author} {\bibfnamefont {S.}~\bibnamefont {Assali}},
  \bibinfo {author} {\bibfnamefont {J.}~\bibnamefont {Nicolas}}, \ and\
  \bibinfo {author} {\bibfnamefont {O.}~\bibnamefont {Moutanabbir}},\ }\href
  {\doibase 10.1088/1361-6641/ab9846} {\bibfield  {journal} {\bibinfo
  {journal} {Semiconductor Science and Technology}\ }\textbf {\bibinfo {volume}
  {35}},\ \bibinfo {pages} {095006} (\bibinfo {year} {2020})}\BibitemShut
  {NoStop}%
\bibitem [{\citenamefont {Wood}\ and\ \citenamefont {Zunger}(1988)}]{Wood1988}%
  \BibitemOpen
  \bibfield  {author} {\bibinfo {author} {\bibfnamefont {D.~M.}\ \bibnamefont
  {Wood}}\ and\ \bibinfo {author} {\bibfnamefont {A.}~\bibnamefont {Zunger}},\
  }\href {\doibase 10.1103/physrevlett.61.1501} {\bibfield  {journal} {\bibinfo
   {journal} {Physical Review Letters}\ }\textbf {\bibinfo {volume} {61}},\
  \bibinfo {pages} {1501} (\bibinfo {year} {1988})}\BibitemShut {NoStop}%
\bibitem [{\citenamefont {Zunger}\ and\ \citenamefont
  {Wood}(1989)}]{Zunger1989}%
  \BibitemOpen
  \bibfield  {author} {\bibinfo {author} {\bibfnamefont {A.}~\bibnamefont
  {Zunger}}\ and\ \bibinfo {author} {\bibfnamefont {D.}~\bibnamefont {Wood}},\
  }\href {\doibase 10.1016/0022-0248(89)90180-2} {\bibfield  {journal}
  {\bibinfo  {journal} {Journal of Crystal Growth}\ }\textbf {\bibinfo {volume}
  {98}},\ \bibinfo {pages} {1} (\bibinfo {year} {1989})}\BibitemShut {NoStop}%
\bibitem [{\citenamefont {Ozoli{\c{n}}{\v{s}}}\ \emph
  {et~al.}(1998{\natexlab{a}})\citenamefont {Ozoli{\c{n}}{\v{s}}},
  \citenamefont {Wolverton},\ and\ \citenamefont {Zunger}}]{Ozolins1998b}%
  \BibitemOpen
  \bibfield  {author} {\bibinfo {author} {\bibfnamefont {V.}~\bibnamefont
  {Ozoli{\c{n}}{\v{s}}}}, \bibinfo {author} {\bibfnamefont {C.}~\bibnamefont
  {Wolverton}}, \ and\ \bibinfo {author} {\bibfnamefont {A.}~\bibnamefont
  {Zunger}},\ }\href {\doibase 10.1103/physrevb.57.4816} {\bibfield  {journal}
  {\bibinfo  {journal} {Physical Review B}\ }\textbf {\bibinfo {volume} {57}},\
  \bibinfo {pages} {4816} (\bibinfo {year} {1998}{\natexlab{a}})}\BibitemShut
  {NoStop}%
\bibitem [{\citenamefont {Hoffman}\ \emph {et~al.}(1989)\citenamefont
  {Hoffman}, \citenamefont {Meyer}, \citenamefont {Wagner}, \citenamefont
  {Bartoli}, \citenamefont {Engelhardt},\ and\ \citenamefont
  {Höchst}}]{Hoffman1989}%
  \BibitemOpen
  \bibfield  {author} {\bibinfo {author} {\bibfnamefont {C.~A.}\ \bibnamefont
  {Hoffman}}, \bibinfo {author} {\bibfnamefont {J.~R.}\ \bibnamefont {Meyer}},
  \bibinfo {author} {\bibfnamefont {R.~J.}\ \bibnamefont {Wagner}}, \bibinfo
  {author} {\bibfnamefont {F.~J.}\ \bibnamefont {Bartoli}}, \bibinfo {author}
  {\bibfnamefont {M.~A.}\ \bibnamefont {Engelhardt}}, \ and\ \bibinfo {author}
  {\bibfnamefont {H.}~\bibnamefont {Höchst}},\ }\href {\doibase
  10.1103/physrevb.40.11693} {\bibfield  {journal} {\bibinfo  {journal}
  {Physical Review B}\ }\textbf {\bibinfo {volume} {40}},\ \bibinfo {pages}
  {11693} (\bibinfo {year} {1989})}\BibitemShut {NoStop}%
\bibitem [{\citenamefont {Calavita}\ \emph {et~al.}(1990)\citenamefont
  {Calavita}, \citenamefont {Engelhart},\ and\ \citenamefont
  {H\"{o}chst}}]{1990PS}%
  \BibitemOpen
  \bibfield  {author} {\bibinfo {author} {\bibfnamefont {E.}~\bibnamefont
  {Calavita}}, \bibinfo {author} {\bibfnamefont {M.~A.}\ \bibnamefont
  {Engelhart}}, \ and\ \bibinfo {author} {\bibnamefont {H\"{o}chst}},\ }in\
  \href {\doibase 10.1016/0022-0248(81)90506-6} {\emph {\bibinfo {booktitle}
  {The Physics of Semiconductors: Proceedings of the 20th International
  Conference}}}\ (\bibinfo {year} {1990})\BibitemShut {NoStop}%
\bibitem [{\citenamefont {Ozoli{\c{n}}{\v{s}}}\ \emph
  {et~al.}(1998{\natexlab{b}})\citenamefont {Ozoli{\c{n}}{\v{s}}},
  \citenamefont {Wolverton},\ and\ \citenamefont {Zunger}}]{Ozolins1998a}%
  \BibitemOpen
  \bibfield  {author} {\bibinfo {author} {\bibfnamefont {V.}~\bibnamefont
  {Ozoli{\c{n}}{\v{s}}}}, \bibinfo {author} {\bibfnamefont {C.}~\bibnamefont
  {Wolverton}}, \ and\ \bibinfo {author} {\bibfnamefont {A.}~\bibnamefont
  {Zunger}},\ }\href {\doibase 10.1063/1.120778} {\bibfield  {journal}
  {\bibinfo  {journal} {Applied Physics Letters}\ }\textbf {\bibinfo {volume}
  {72}},\ \bibinfo {pages} {427} (\bibinfo {year}
  {1998}{\natexlab{b}})}\BibitemShut {NoStop}%
\bibitem [{\citenamefont {Asano}\ \emph {et~al.}(2014)\citenamefont {Asano},
  \citenamefont {Kidowaki}, \citenamefont {Kurosawa}, \citenamefont {Taoka},
  \citenamefont {Nakatsuka},\ and\ \citenamefont {Zaima}}]{Asano2014}%
  \BibitemOpen
  \bibfield  {author} {\bibinfo {author} {\bibfnamefont {T.}~\bibnamefont
  {Asano}}, \bibinfo {author} {\bibfnamefont {S.}~\bibnamefont {Kidowaki}},
  \bibinfo {author} {\bibfnamefont {M.}~\bibnamefont {Kurosawa}}, \bibinfo
  {author} {\bibfnamefont {N.}~\bibnamefont {Taoka}}, \bibinfo {author}
  {\bibfnamefont {O.}~\bibnamefont {Nakatsuka}}, \ and\ \bibinfo {author}
  {\bibfnamefont {S.}~\bibnamefont {Zaima}},\ }\href {\doibase
  10.1016/j.tsf.2013.10.087} {\bibfield  {journal} {\bibinfo  {journal} {Thin
  Solid Films}\ }\textbf {\bibinfo {volume} {557}},\ \bibinfo {pages} {159}
  (\bibinfo {year} {2014})}\BibitemShut {NoStop}%
\bibitem [{\citenamefont {O'Donnell}\ \emph {et~al.}(2020)\citenamefont
  {O'Donnell}, \citenamefont {Sanchez-Soares}, \citenamefont {Broderick},\ and\
  \citenamefont {Greer}}]{BandGapPaper}%
  \BibitemOpen
  \bibfield  {author} {\bibinfo {author} {\bibfnamefont {C.}~\bibnamefont
  {O'Donnell}}, \bibinfo {author} {\bibfnamefont {A.}~\bibnamefont
  {Sanchez-Soares}}, \bibinfo {author} {\bibfnamefont {C.~A.}\ \bibnamefont
  {Broderick}}, \ and\ \bibinfo {author} {\bibfnamefont {J.~C.}\ \bibnamefont
  {Greer}},\ }\href {https://arxiv.org/abs/2012.11311} {\bibfield  {journal}
  {\bibinfo  {journal} {arXiv:2012.11311}\ } (\bibinfo {year}
  {2020})}\BibitemShut {NoStop}%
\end{thebibliography}
\end{document}